\documentclass[10pt,conference]{IEEEtran}
\usepackage{cite}
\usepackage{amsmath,amssymb,amsfonts}
\usepackage{algorithmic}
\usepackage{graphicx}
\usepackage{textcomp}
\usepackage{xcolor}
\usepackage[hyphens]{url}
\usepackage{fancyhdr}
\usepackage{hyperref}

\usepackage{array}
\usepackage{dblfloatfix}
\usepackage{placeins}
\usepackage{multirow}
\usepackage{threeparttable}
\usepackage{algorithm}
\usepackage{algorithmic}
\usepackage[numbers,sort&compress]{natbib} 
\usepackage{pifont}
\ifCLASSOPTIONcompsoc
\usepackage[caption=false, font=normalsize, labelfont=sf, textfont=sf]{subfig}
\else
\usepackage[caption=false, font=footnotesize]{subfig}
\fi
\newcommand{\hpcayear}{2024}

\newcommand{\mv}[1]{{\color{black}#1}}
\newcommand{\mvn}[1]{{\color{black}#1}}
\newcommand{\vs}[1]{{\color{black}#1}}
\newcommand{\hpcasubmissionnumber}{NaN}
\title{BitWave: Exploiting Column-Based Bit-Level Sparsity for Deep Learning Acceleration}

\def\hpcacameraready{} 

\newcommand\hpcaauthors{Man Shi$^\ast$, Vikram Jain$^\ast$, Antony Joseph$^\ddagger$, Maurice Meijer$^\ddagger$, Marian Verhelst$^\ast$}
\newcommand\hpcaaffiliation{Department of Electrical Engineering - MICAS, KU Leuven, Leuven, Belgium $^\ast$, \\ NXP Semiconductor, Belgium$^\ddagger$}
\newcommand\hpcaemail{Email: \{man.shi, marian.verhelst\}@kuleuven.be, \\ vikramj@berkeley.edu, \{antony.joseph, maurice.meijer\}@nxp.be}

\author{
  \ifdefined\hpcacameraready
    \IEEEauthorblockN{\hpcaauthors{}}
      \IEEEauthorblockA{
        \hpcaaffiliation{} \\
        \hpcaemail{}
      }
  \else
    \IEEEauthorblockN{\normalsize{HPCA \hpcayear{} Submission
      \textbf{\#\hpcasubmissionnumber{}}} \\
      \IEEEauthorblockA{
        Confidential Draft \\
        Do NOT Distribute!!
      }
    }
  \fi 
}

\fancypagestyle{camerareadyfirstpage}{%
  \fancyhead{}
  
  \fancyhead[C]{
    \ifdefined\aeopen
    \parbox[][12mm][t]{13.5cm}{\hpcayear{} IEEE International Symposium on High-Performance Computer Architecture (HPCA)}    
    \else
      \ifdefined\aereviewed
      \parbox[][12mm][t]{13.5cm}{\hpcayear{} IEEE International Symposium on High-Performance Computer Architecture (HPCA)}
      \else
      \ifdefined\aereproduced
      \parbox[][12mm][t]{13.5cm}{\hpcayear{} IEEE International Symposium on High-Performance Computer Architecture (HPCA)}
      \else
      \parbox[][0mm][t]{13.5cm}{\hpcayear{} IEEE International Symposium on High-Performance Computer Architecture (HPCA)}
    \fi 
    \fi 
    \fi 
    \ifdefined\aeopen 
      \includegraphics[width=12mm,height=12mm]{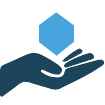}
    \fi 
    \ifdefined\aereviewed
      \includegraphics[width=12mm,height=12mm]{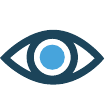}
    \fi 
    \ifdefined\aereproduced
      \includegraphics[width=12mm,height=12mm]{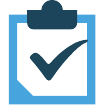}
    \fi
  }
  \fancyfoot[C]{}
}
\begin{document}
\maketitle

\ifdefined\hpcacameraready 
  \thispagestyle{camerareadyfirstpage}
  \pagestyle{empty}
\else
  \thispagestyle{plain}
  \pagestyle{plain}
\fi

\newcommand{\hpcaheight}{0mm}
\ifdefined\eaopen
\renewcommand{\hpcaheight}{12mm}
\fi

\begin{abstract}
Bit-serial computation facilitates bit-wise sequential data processing, offering numerous benefits, such as a reduced area footprint and dynamically-adaptive computational precision. It has emerged as a prominent approach, particularly in leveraging bit-level sparsity in Deep Neural Networks (DNNs). However, existing bit-serial accelerators exploit bit-level sparsity to reduce computations by skipping zero bits, but they suffer from inefficient memory accesses due to the irregular indices of the non-zero bits.

As memory accesses typically are the dominant contributor to DNN accelerator performance, this paper introduces a novel computing approach called "bit-column-serial" and a compatible architecture design named "\emph{BitWave}." \emph{BitWave} harnesses the advantages of the "bit-column-serial" approach, leveraging structured bit-level sparsity in combination with dynamic dataflow techniques. This achieves a reduction in computations and memory footprints through redundant computation skipping and weight compression. \emph{BitWave} is able to mitigate the performance drop or the need for retraining that is typically associated with sparsity-enhancing techniques using a post-training optimization involving selected weight bit-flips.
Empirical studies conducted on four deep-learning benchmarks demonstrate the achievements of \emph{BitWave}: (1) Maximally realize $13.25\times$ higher speedup, $7.71\times$ efficiency compared to state-of-the-art sparsity-aware accelerators. (2) Occupying $1.138$ $mm^2$ area and consuming $17.56$ mW power in 16nm FinFet process node.


\end{abstract}

\section{Introduction}
Deep neural networks (DNNs) are constructed with ever-increasing model size and complexity to enhance accuracy. As a result, the performance of deep learning accelerators must at the same time align with the growing performance demands and the increased pressure on power-efficient deployment. Particularly in embedded edge devices like digital watches, smartphones, or autonomous vehicles, the latter is driven by the limitations imposed by hardware resources, power budget, and/or cost. Therefore, there is a strong need to further enhance the efficiency of these accelerators to cater to both high-performance and power-efficient use cases.

At the algorithmic level, several techniques aim at reducing the size of deep learning models while maintaining performance and accuracy. Quantization \cite{han2015deep, liang2021pruning, yang2019quantization, tang2022mixed, gholami2022survey}, one of the most popular techniques, can effectively compress the size and runtime overhead of a NN model by reducing the bit-width of the network operand. For instance, quantization-aware training can reduce the bit-width significantly below 8-bit.
Yet, this comes with the drawback of requiring time-consuming and resource-intensive retraining, necessitating access to the original training dataset or special training strategies \cite{choi2016towards}. Unfortunately, people, especially in the industry, are reluctant to share training datasets or modify the training method due to privacy concerns and sharing IP constraints. The alternative quantization approach using post-training quantization (PTQ) also allows precision scaling but suffers several tradeoff between precision and model accuracy \cite{xiao2023smoothquant, yao2022zeroquant, wei2023qdrop}.
\begin{figure}[tb]
\centering
\includegraphics[width=0.98\linewidth]{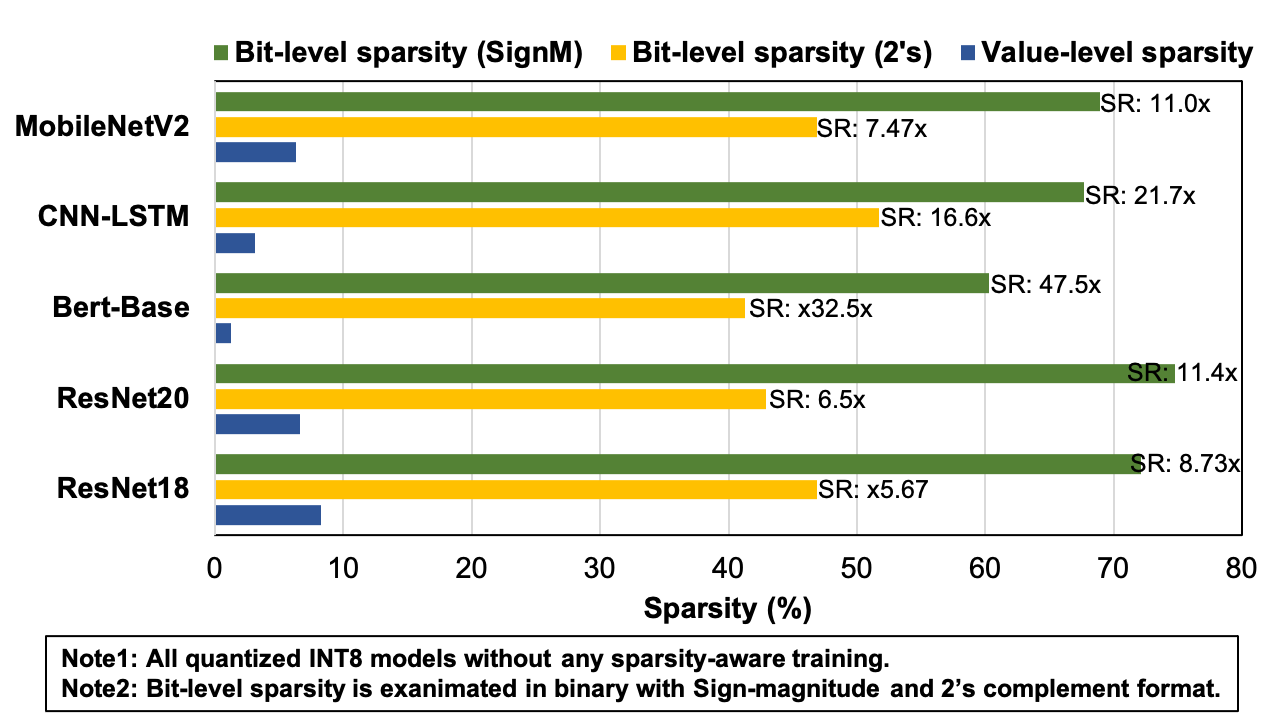}
\caption{Comparison of weight value sparsity and bit sparsity in Int8 DNNs. SR indicates the sparsity ratio between weight bit-level sparsity and weight value-level sparsity.} 
\label{fig:f1} 
\vspace{-0.5cm}
\end{figure}
Besides using quantization for model compression, numerous studies \cite{bitlet, li2022ristretto, 10071080, gondimalla2019sparten} have leveraged sparsity to enhance the efficiency of DNN accelerators further. 
However, sparsity in NN models shows strong randomness, whether in an individual layer or the entire model. Activation sparsity, for example, has a higher likelihood in the presence of non-linear activation functions such as ReLU or PReLU, while other activation functions like Sigmoid, LeakyReLU, GeLU, or Tanh tend to exhibit limited activation sparsity. Conversely, sparsity is generally scarce in the quantized weights of networks trained without sparsity-enhancing loss functions.
Some approaches, \mv{therefore,} implement laborious sparse (re)training techniques to induce more sparsity \mv{and} pruning \cite{zhu2017prune, wen2016learning, Xia_2023_CVPR, liu2022s2ta}. However, \mv{such value-based sparsity exploitation comes with significant challenge:} 
On the one hand, regardless of the pruning methodology employed, considerable time is spent exploring the desired balance between accuracy and model size. On the other hand, the limited and irregular sparsity that arises without (re)training
leads to \mv{inefficient} memory accesses or data decoding in the memory interface, posing a bottleneck in existing spatially-parallel accelerators \cite{parashar2017scnn, zhang2016cambricon, li2022ristretto}. 


%
%


To address the aforementioned challenges, state-of-the-art (SotA) solutions in the sparsity domain leverage "bit-level sparsity (BLS)". This form of sparsity focuses on the "zero bits" within the operand word. As shown in Figure~\ref{fig:f1}, the value-based sparsity on a series of popular 8-bit quantized networks is about one order of magnitude lower than the networks' BLS \mvn{in 2's complement notation, bringing} \vs{a potential computational speedup \mvn{of} 5.67$\times-32.5\times$. 
\mvn{Furthermore,} we observe that sign-magnitude-based binaries exhibit even higher sparsity, ranging from 8.73$\times-47.5\times$ compared to value sparsity.} Instead of eliminating computations caused by zero values, selectively skipping zero bits in operands presents a promising avenue for achieving more energy- and throughput-efficient DNN deployment. Yet, significant savings are only possible if zero bits can be skipped efficiently without introducing inefficiencies stemming from the sparsity-induced computing irregularity.

To this end, several bit-serial (BS) accelerators \cite{delmas2019bit, bitlet, sharify2019laconic, albericio2017bit, yang2021fusekna} proposed to leverage the abundant BLS to varying degrees. To overcome the irregularity of BLS, BitCluster \cite{li2022bitcluster}, BitPruner \cite{zhao2020bitpruner}, and Bit-balance \cite{sun2023bit} adopt hardware-software co-design to induce structured BLS through the training. Yet, this again faces the same training downsides as quantization-aware training. Bit-Tactical \cite{delmas2019bit}, Fused kernel \cite{yang2021fusekna}, and Bitlet \cite{bitlet} designed complex online hardware bit-schedulers to try to balance the workload at runtime to utilize the BLS maximally. 
However, these approaches have one significant weakness, as they design sophisticated hardware modules for skipping ineffective computations caused by the highly irregular zero bits in the common two’s complement operand format, but fail to optimize the costly memory accesses. Additionally, these BS accelerators suffer significantly from PE array under-utilization, since BS operation requires strongly increased parallelism of the PE array to achieve the same throughput compared to bit-parallel PE. The resulting losses in overall system efficiency when running actual workloads have received limited attention in previous related works. 

To tackle these problems, this work proposes \emph{BitWave}, combining "Bit-Column-Serial Computation" (BCSeC) with an aligned hardware-optimized dataflow. BCSeC exploits the sign-magnitude data representation to enable efficient column-based BLS of DNN weights without involving any (re)training. The BCSeC-optimized dataflow, moreover maximizes the resulting PE array utilization in a hardware-friendly manner.
The major highlights of this work are as follows:
\begin{enumerate}
	\item \mv{Exploiting inherent structured and more abundant BLS in sign-magnitude weight representations to enable increased weight compression and hardware-efficient bit-column-serial computations} (Section \ref{sec:b1.1} - \ref{sec:b1.3}). 
        \item Mitigating the performance drop or the need for any retraining typically associated with sparsity-enhancing techniques by selectively flipping bits in a single-shot post-training optimization step 
        (\ref{sec:b1.4}).
        \item Presenting an architecture aligned DNN spatial and temporal dataflow, \emph{BitWave}, meticulously tailored to harness the advantages of the BCSeC 
        (Section \ref{sec:architeture}).
	\item Comparing \emph{BitWave} against SotA DNN accelerators, demonstrating up to $7.71\times$ energy efficiency and $13.25\times$ performance improvement in Section \ref{sec:Evaluation}.

\end{enumerate}
\begin{figure}[tb]
\centering
\includegraphics[width=0.8\linewidth]{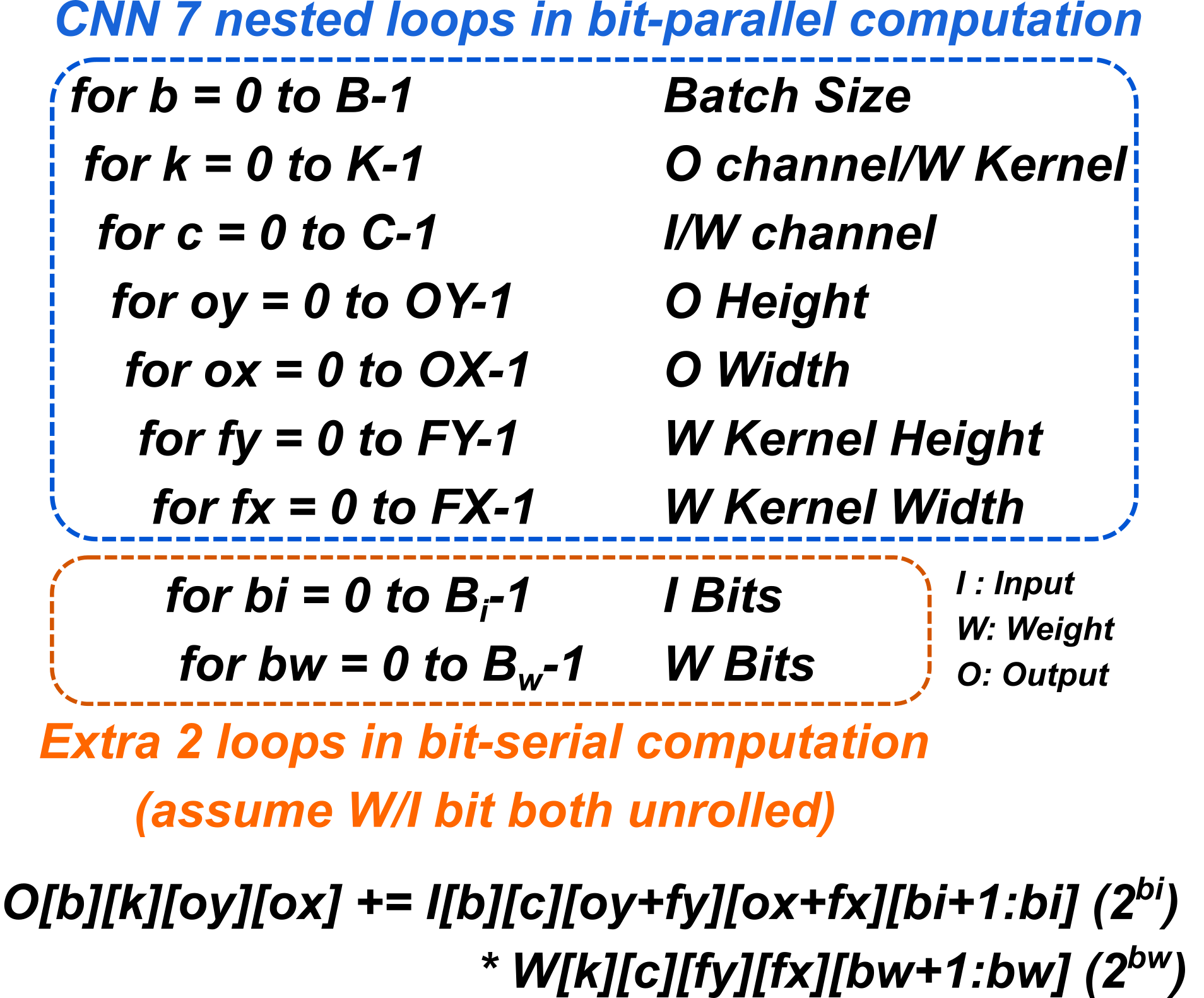}
\caption{For-loops of example CNN workload in bit-parallel computation and bit-serial computation} 
\label{fig:loop}   
\vspace{-0.3cm}
\end{figure}

\section{Background and Motivation}\label{sec:s2}

\subsection{Dataflow}
\label{sec:df}
The representation of a typical NN layer often involves nested for-loops. Figure~\ref{fig:loop} illustrates the for-loops of the CNN layer workload. These loops encompass the dimensions of both the output feature map (OX, OY, K, B) and the input feature map (IX, IY, C), as well as the shape of weight kernels (FX, FY, C) in bit-parallel computation.
By reordering and splitting these loops, the NN layer can be efficiently processed in hardware, offering a wide range of dataflow strategies (execution orders). These dataflows involve diverse parallelization choices (spatial unrolling, SU) and loop tiling/ordering schemes (temporal unrolling, TU). SU  focuses on parallelizing the layer's loop dimensions within a single clock cycle, enabling the simultaneous processing of multiple data elements. It determines how the innermost loop dimensions are expanded across the MAC (multiply-accumulate) array. The suffix '\textbf{u}' is used to denote this parallelism. For instance, $OX_u, OY_u, C_u, K_u, FX_u, FY_u$ represent the number of data elements required from each loop dimension per clock cycle. On the other hand, TU refers to the unrolling of the remaining loop dimensions in a temporal manner. The specific ordering and tiling methods employed for these loops impact the data layout and memory access patterns within the memory hierarchy when executing different NN layers.


\subsection{Bit-serial computation}
\label{sec:bitserial}    
\begin{figure}[tb]
\centering
\includegraphics[width=1\linewidth]{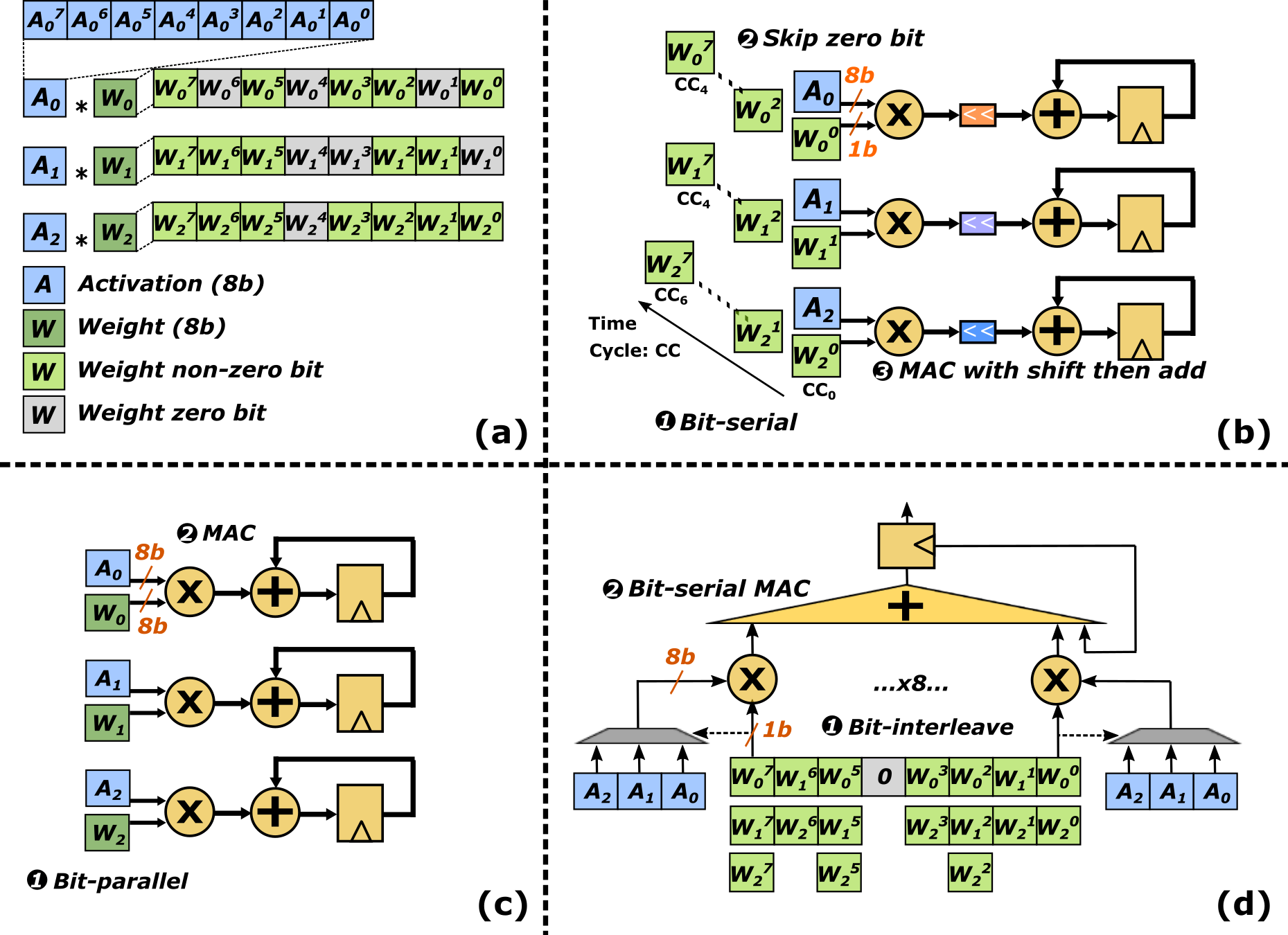}
\caption{\vs{The comparison among three types of PE. (a) Computation example. (b) Bit-serial PE. (c) Bit-parallel PE. (d) Bit-interleaved PE.}} 
\label{fig:pe}   
\vspace{-0.3cm}
\end{figure}

It is crucial to note that MAC operation in the inner loop holds two additional nested for-loops in BS computation, specifically across the bits of the input $B_i$ and the weight $B_w$. Similar to the SU and TU techniques applied to the output for-loops, one can also decide to spatially unroll along $B_i$ and $B_w$ (as done in bit-parallel architectures, Figure~\ref{fig:pe} (c)), or to unroll one or both of them (partially) \vs{temporally}, as done in BS architectures (Figure~\ref{fig:pe} (b)).
BS computation hence 
involves slicing multi-bit data into individual bits and transforming multiplications into sequential additions. While bit-parallel computing elements are typically blind to BLS, bit-serial PE's can more easily skip ineffective additional zero bits. This approach has also gained popularity for accelerating computations in asymmetric and varying-precision DNNs.
Yet, a limitation of BS computation is the inefficiency in memory accesses caused by the \vs{irregularity of} zero bits. To exploit the maximum potential of the bit sparsity, existing BS accelerators (Figure~\ref{fig:pe} (b)) tend to skip the zero bits to maximally reduce redundant computations, assisted with complex synchronization mechanisms to carefully implement the bit-significance alignment \cite{delmas2019bit, yang2021fusekna}. \vs{Also, Bitlet \cite{bitlet} can leverage \mvn{such} bit-level sparsity (Figure~\ref{fig:pe} (d)). However, it requires expensive online preprocessing, to decide at runtime on \mvn{which computations to skip and how to} interleave non-zero bits \mvn{for} parallel computation. As such, Bitlet necessitates extensive \mvn{runtime processing} to extract the \mvn{indices} of non-zero weight bits, significantly increasing memory overhead and hampering efficiency.}
Most of implementations, however, fail to optimize the inefficient memory accesses, as the irregularity of the zero-bit locations prevents effective data compression and efficient memory utilization. The performance and energy efficiency of BS accelerators are further degraded by increased under-utilization of the PE arrays 
compared to bit-parallel architectures. As illustrated in Figure~\ref{fig:loop}, BS computation temporally unrolls the $B_w$ and/or $B_i$ loops. Consequently, to maintain computational throughput, the spatial unrolling dimensions of the other for-loops in the stack will have to be increased $B_{i,u}*B_{w,u}$-times compared to bit-parallel processing. However, these increased unrolling dimensions of $C$, $K$, $OX$, etc., hinder the mapping of a wide variety of layers (with different loop dimensions) at high spatial utilization. Several studies \cite{10129330, 9076333, 9407116, du202328nm, 9950755} have highlighted the resulting issue of severe under-utilization in larger PE arrays. 

Motivated by the challenges associated with traditional BS computation, we introduce a novel approach called Bit-Column-Serial Computation (BCSeC) with optimized dataflow. This approach aims to overcome the limitations of BS computation by effectively leveraging BLS while optimizing computation and memory footprints, without modifications to the network training. 




\section{Bit-Column Serial Computation}
\label{sec:BitWave}
\begin{figure*}[tb]
\centering
\includegraphics[width=0.9\linewidth]{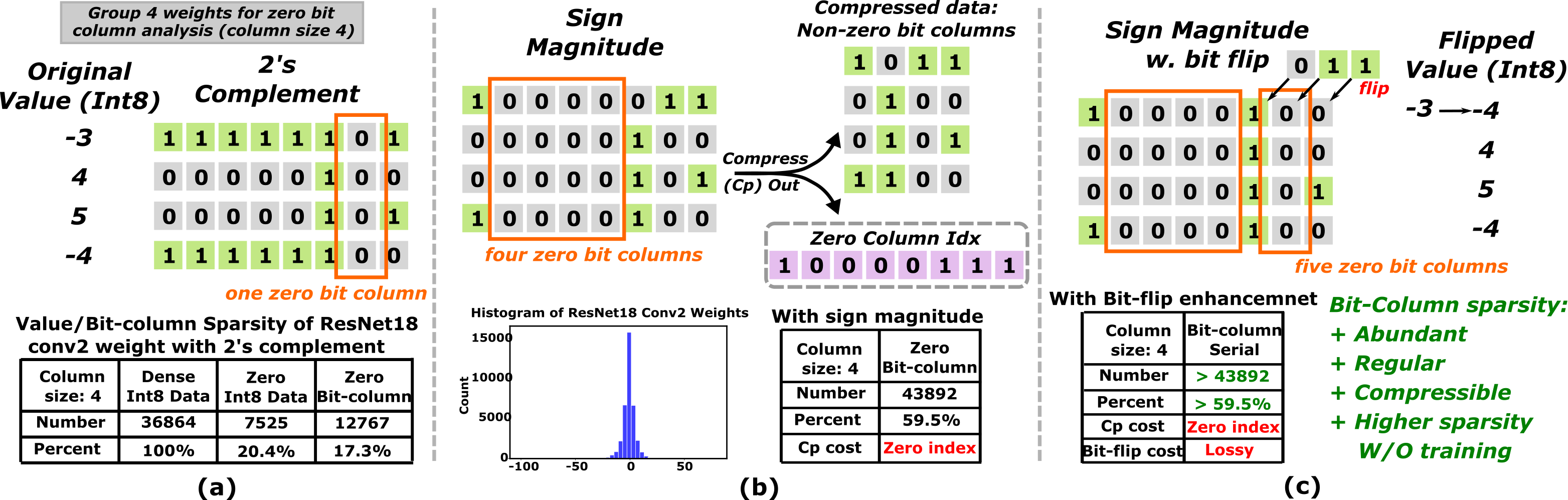}
\caption{ \mvn{Bitgroup visualization with a group size of $G=4$:} \vs{(a) Bit-column sparsity with two's complement representation. (b) Bit-column compression with sign-magnitude representation. (c) Sign-magnitude-based bit-column sparsity with Bit-Flip enhancement.}} 
\label{fig:bit}   
\vspace{-0.4cm}
\end{figure*}

\begin{figure}[tb]
\centering
\includegraphics[width=1\linewidth]{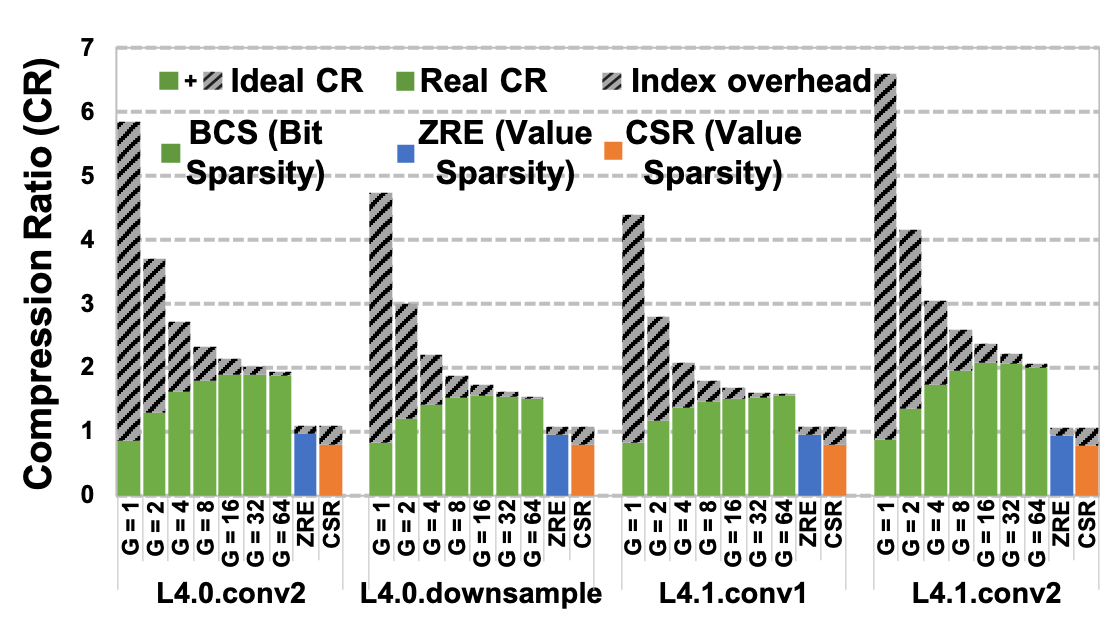}
\caption{\mvn{Compression Rate (CR) under Zero Run Length Encoding (ZRE), Compressed Sparse Row (CSR) to value sparsity and BCS-compression for ResNet18's last 4 conv layers. BCS-compression for varying group size (G). All bars visualize both the ideal compression ratio without index overheads, as well as the real CR with the consideration of index cost.}
$CR = Size(Data_{orig.}) / Size(Data_{compre})$.}  
\label{fig:bcs}   
\vspace{-0.4cm}
\end{figure}
We propose Bit-Column-Serial Computation  (BCSeC) to leverage the abundant BLS in DNNs without (re)training involvement. The BCSeC is achieved by effectively skipping the computation caused by patterned BLS, which we call Bit-Column Sparsity (BCS). Therefore, we first introduce the BCS based on 2's complement formatted weight operands.
Next, 
the extended BCS is explored in operands using sign-magnitude representation, and subsequently we quantify its benefits for weight compression. We finally present a single-shot post-training optimization step, denoted by Bit-Flip, which further enhances BCS by selectively flipping bits in binary weights.

\vspace{-0.2cm}
\subsection{\textbf{Bit-column Sparsity}}\label{sec:b1.1}
Instead of representing individual weights separately, BCS groups multiple weights together and examines their BLS collectively. In Figure~\ref{fig:bit}(a), we illustrate this approach using four Int8 values in the commonly-used 2's complement binary representation. Apart from value-level sparsity requiring all bits within one Int8 data item to be zero, BCS focuses on zeroes occurring at the same bit significance across all data elements of the group. In the example, such a zero-bit column exists for the $LSB+1$-position across the four grouped Int8 data. Like traditional BS computation, the patterned zero-bit column can be temporarily skipped by only streaming non-zero-bit columns into PE to save energy and reduce latency. Moreover, grouping multiple data elements mitigates the indexing overhead and control hardware across more operations, increasing efficiency compared to traditional BLS. 

However, BCS does not fulfill its potential in combination with 2's complement binary representation. The table in Figure~\ref{fig:bit}(a) indicates \vs{the count of zero-bit columns in the layer conv2 of ResNet18, assuming groups of 4} weight elements from consecutive input channels of one kernel. 
While this layer possesses 20$\%$ zero-values, it only achieves 17$\%$ sparsity due to zero-columns. This discrepancy stems from the presence of small negative values in the parameters, represented by several leading 1's in the 2's complement representation, significantly reducing the BCS. 
To enhance BCS, we switch the binary data representation to the sign-magnitude format. 

\subsection{\textbf{Bit-column Sparsity With Sign-magnitude}}\label{sec:b1.2}
It is widely recognized that NN weights often exhibit non-uniform distributions with a high frequency of small or zero values \cite{han2015deep, an202329}, as also apparent in the histogram of ResNet18 conv2 weights (Figure~\ref{fig:bit}(b)). These dominant negative weights with small absolute values can contribute significantly to achieving more BCS when utilizing the sign-magnitude (SM) binary representation. In the running example of Figure~\ref{fig:bit}(b), the adoption of the SM representation introduces four zero columns. When assessing the impact on the actual workload of ResNet18 conv2, simply switching to the SM format increases the bit column sparsity to 59$\%$, or a $3.4\times$ improvement in the bit column-wise sparsity compared to the 2's complement format. 

\subsection{\textbf{Sign-magnitude-based BCS With Compression}}
\label{sec:b1.3}
BCS not only allows to eliminate computations but also to reduce the memory footprints. This stems from the fact that BCS enables the compression of zero-columns within grouped data words, granting for storing only the non-zero columns with their index information. As the bit significance information is shared by all data elements in the group, it can be specified through a zero-column index, as illustrated in Figure~\ref{fig:bit}(b). This index has the same length as the operand precision with each bit indicating the occurrence of a non-zero (1) or zero (0) column at that bit position.  
In contrast to other BLS-based approaches, which due to irregularity fail to achieve significant compression rates or reduced memory accesses, 
BCS is a lossless compression technique, which provides regularity at the bit column level. 

To evaluate our proposed technique, we explore the trade-off between different grouped sizes (column size, G) and the index overhead in BCS. 
Larger column sizes would result in smaller index overheads, as one index bit can be shared across more bits. Figure~\ref{fig:bcs} 
shows the total compression ratio (CR) 
when analyzing BCS on the last \vs{four conv layers of ResNet18 (accounting for $\geq$ 50$\%$ of its network weights)} with varying column sizes, ranging from 1, 2, 4, 8, 16, 32 to 64. The highest CR occurs when the column size is 1, as it is equivalent to the number of zero bits within the weights. Yet, the significant cost of the index offsets the compression benefits. As the column size increases, CR gradually reduces due to less co-occurring zeros across all data elements in the larger groups. \vs{Additionally, we compared the CR using common value-sparsity-based compression techniques like Zero Run Length Encoding (ZRE) and Compressed Sparse Row (CSR) \mvn{formatting} \cite{wu2023sparseloop}. The results \mvn{show that} BCS-based compression consistently outperforms other value-sparsity techniques, especially \mvn{under} low-value sparsity scenarios. }
It is important to note that different networks, or even layers within the same network, may have varied optimal column sizes. Therefore, the \emph{BitWave} realization will support layer-wise tunable column sizes of 8, 16, and 32, which is able to cover most cases and can explore the optimal trade-off between CR and index overheads in our experiments.

\begin{figure*}[tb]
\centering
\includegraphics[width=1\linewidth]{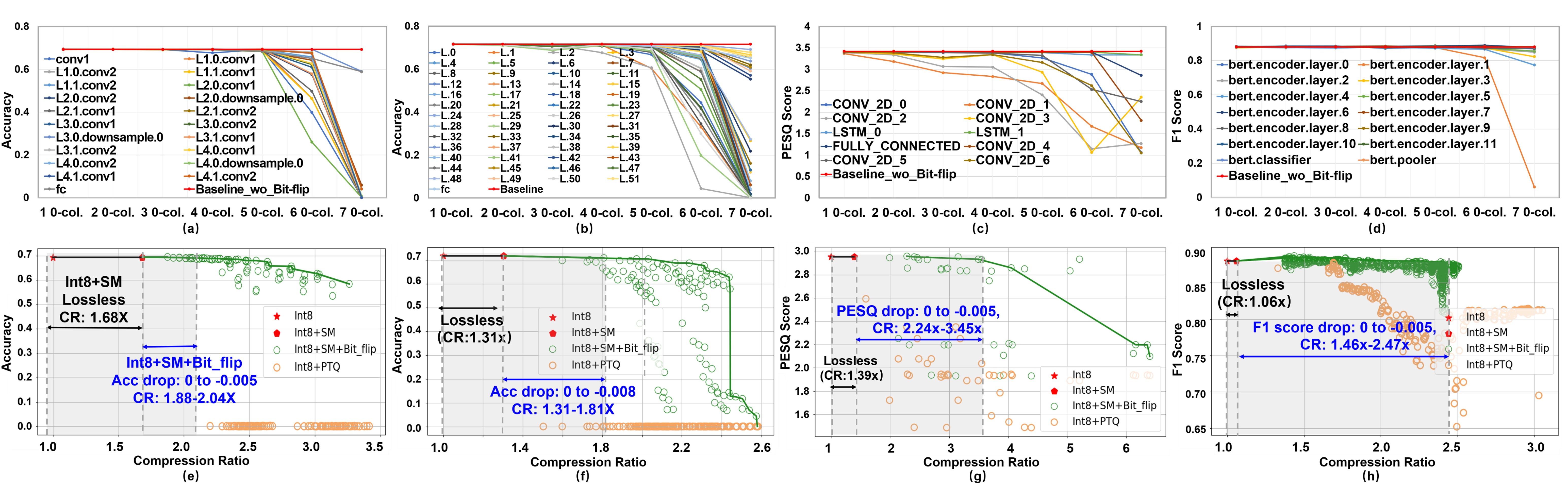}

\caption{(a), (b), (c) (d) show Resnet18/MobileNet/CNN-LSTM/Bert-Base layer-wise weight flipping sensitivity, respectively. (e) ResNet18 CR v.s. accuracy, applying 4 to 7 zero columns on the L.4.0, L4.1, and fc ($70\%$ weights),  remaining layers untouched; (f), MobileNetv2 CR v.s. accuracy, applying 4 to 7 zero columns on L.47, L.48. L50, L.51 and fc ($70\%$ weights), remaining layers untouched; (g), CNN-LSTM CR v.s. PESQ score, applying 4 to 7 zero columns on LSTM.0 and LSTM.1 ($80\%$ weights), remaining layers untouched. (h), Bert-Base CR v.s. F1 score, applying Bit-Flip on sensitive layers (Layer.1, Layer.2, Layer.3) from 0 to 4 zero-columns and flipping other layers from 4 to 7 zero-columns.}
\label{fig:wdis}   
\vspace{-0.4cm}
\end{figure*}

\subsection{\textbf{Bit Flip Enhancement}}
\label{sec:b1.4}
To further increase the benefits from BCS, we propose a fine-grained weight bit adjustment strategy called "Bit-Flip". In contrast to the approach discussed in previous sections, this approach is no longer lossless but will aim to manipulate the original weight vector, to achieve an increased number of zero-bit columns.
The Bit-Flip algorithm operates on the original weight vector and iteratively modifies individual weights. It can find the closest weight vector (measured by root mean square (RMS) value) that satisfies a specified constraint on the desired number of zero-bit columns. This objective ensures that an equal number of zero-bit columns can be manipulated across the entire layer, leading to a balanced workload during parallel execution.
Therefore, the Bit-Flip algorithm aims to minimize the Euclidean Distance between the modified and original weight vectors while ensuring the desired sparsity level. For instance, in Figure~\ref{fig:bit}(c), if we target five zero-bit columns, the binary (-3) tunes to $8’b1000100 (-4)$ with a vector distance of 1, which will be shown to be tolerable due to the redundancy in DNNs \cite{sun2023bit}.

\begin{algorithm}
    \small
    \renewcommand{\algorithmicrequire}{\textbf{Input:}}
    \renewcommand{\algorithmicensure}{\textbf{Output:}}
    \caption{Greedy Search Bit-Flip Strategy}
    \label{alg3} 
    \begin{algorithmic}[1] 
    \REQUIRE Network: M, Initial Strategy: S, Minimal Accuracy: macc, Dataste: D, Inference Function: Inference
    \STATE M $=$ BitFlip(M,S)
    \WHILE{True}
    \STATE $S_{tmp} = S$
    \STATE bacc $=0$
    \FOR{layer in M}
    \FOR{$gs$ in $8,16,32$}
        \STATE Get number of zero column: z $= S_{tmp}[layer][gs]$
        \STATE Update $S_{tmp}$: $S_{tmp}[layer][gs]=z+1$
        \STATE M $=$ BitFlip(M,$S_{tmp}$)
        \STATE accuracy $=$ Inference(M, D)
        \IF{accuracy $\geq$ bacc}
        \STATE bacc $=$ accuracy
        \STATE next move = (layer, gs, z+1)
        \ENDIF 
    \ENDFOR
    \STATE $S_{tmp} = S$
    \ENDFOR
    \IF{bacc $\leq$ macc}
    \STATE Break
    \ENDIF
    \STATE Update strategy $S$ with next move
    \ENDWHILE 
    \ENSURE Strategy $S$
    \end{algorithmic} 
\end{algorithm}

To implement the Bit-Flip strategy, we adopt a layer-wise approach to tailor the sparsity level. The steps involved are: 1) Layer-level analysis: We begin by flipping the weights of one layer at a time and monitor the evolution of the resulting number of zero columns relative to model accuracy. This helps us to understand each layer's sensitivity to weight modifications. Our findings illustrated in Figure~\ref{fig:wdis}(a), with ResNet18, reveal that most layers exhibit negligible accuracy degradation when the entire layer is forced to have less than four zero columns. We also observe that early layers (weight-light layers) in ResNet18 tend to be more sensitive than later layers (weight-heavy layers). Similar trends are seen in both CNN-LSTM \cite{cnnlstm} and MobileNetV2 \cite{sandler2018mobilenetv2}. Although, The weights size of each layer of Bert-Base \cite{devlin2019bertpretrainingdeepbidirectional} are equal. We still find some layers especially sensitive, for instance, the bert.encoder.layer.1 as shown in Figure~\ref{fig:wdis}(d). 2) Network-wide optimization: Building upon the insights gained from the layer-level analysis, we selectively flip the weight-flip-non-sensitive layers to more zero columns (4-7), while for weight-flip-sensitive layers to less zero columns (1-4) or leave untouched, ensuring maximization of the compression ratio and minimization of accuracy degradation. 


Employing multi-objective optimization, we search for Pareto points that offer a favorable trade-off between the number of zero columns for each flipped layer and the accuracy. The outcome of this step is a Pareto front that represents the weight configurations to achieve the optimal balance between compression ratio and accuracy. We examine BCS with the varied weight group sizes (8, 16, 32) and different desired numbers of zero columns on the flipped layers. As demonstrated in Algorithm 1, it can return the strategy for all layers that have reached the potential largest number of zero-column with the pre-defined minimal accuracy constraint.  
Figure~\ref{fig:wdis}(e) showcases the obtained Pareto points for ResNet18, demonstrating a compression ratio of $2.04\times$ with less than $0.5\%$ accuracy drop. Notably, CNN-LSTM achieves a compression ratio of $3.45\times$ with near $0.5\%$ PESQ (Perceptual Evaluation of Speech Quality \cite{rix2001perceptual}) drop compared with the base Int8 model. The initial strategy (flipping L.47, L.48. L50, L.51, and FC layer to have at least four zero-bit columns in each weight group) of MobileNetv2 already comes at a $0.8\%$ accuracy drop. However, the compression ratio can reach $1.81\times$, achieving considerable energy-efficiency improvement for this more compact network. Although Bert-Base has a limited number of zero columns in the original Int8 model, it can realize a $1.46\times$ compression ratio without any accuracy drop through Bit-Flip and is capable of achieving high to $2.47\times$ CR with less than $0.5\%$ F1 score degradation.

The results obtained by BCS and the Bit-Flip strategy should, of course, be compared to traditional PTQ \vs{to \mvn{benchmark BCS against simply using a reduced bit width of the same weights}.} 
Figure~\ref{fig:wdis} (e)-(f), therefore, compares three cases reducing the bit-width of parameters to less than 8 bits across four DNNs: Int8+PTQ, Int8+SM, and Int8+SM+Bit-Flip. In Int8+PTQ, we apply PTQ on the original Int8 model by quantizing the weights of each layer to less bits and achieving the same CR compared with the other two methods, \mvn{assessing the accuracy impact in function of the achieved} CR. The lossless Int8+SM compression already outperforms Int8+PTQ in terms of accuracy for the different DNNs.

\mvn{From the results,} the Int8+Bit-Flip+SM approach \mvn{drastically} increases the CR with a negligible accuracy drop. 
\mvn{While} PTQ primarily focuses on reducing the precision of the entire data representation by cutting \mvn{the same} MSB or LSB bit positions \mvn{across a whole tensor}, BCS and Bit-Flip strategies provide more flexibility in achieving sparsity, i.e., \mvn{on the fine granularity of the group,} zero bits can appear \mvn{at any bit position} without significantly affecting the overall binary representation range at the layer or network level. 
This differs from PTQ's reduction in the dynamic range \mvn{across the coarse granularity groups}, which can have a relatively large impact on accuracy, especially when aggressively quantizing operands to less than 8 bits. 
\vs{Even more advanced non-uniform quantization techniques, such as Mr.BiQ [3],}
\mvn{suffer from the need for high precision activations (32 bits), an 8-bit constraint on the first and last layer weights, and/or a need for access to parts of the dataset for fine-tuning or calibration. In contrast, our BCS method can compress weights at a fine group granularity and in a completely "data-free" way, achieving a 2-3.5x compression ratio with only a 0.5\% accuracy drop.}


In summary, the structured BCS can be employed in BCSeC for \vs{simultaneous} compression and computations elimination. However, as discussed in Section \ref{sec:s2}, BCSeC still faces the challenge of under-utilizing the PE array due to its serial nature. To overcome this limitation and maximize the benefits of BCSeC, Section \ref{sec:architeture} will optimize the dataflow of BCSeC and present our proposed architecture.

\begin{figure}[tb]
\centering
\includegraphics[width=0.65\linewidth]{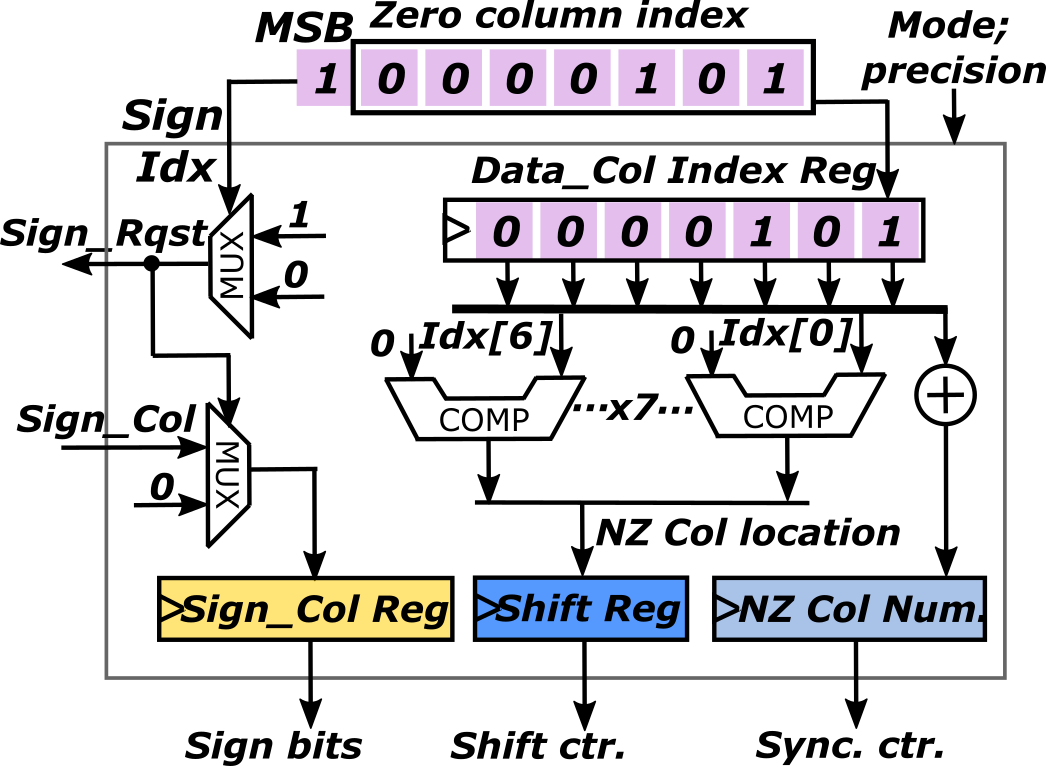}
\caption{\vs{Zero-column Index Parser}} 
\label{fig:idx}  
\vspace{-0.5cm}
\end{figure}

\section{BitWave Architecture Design}\label{sec:architeture}
To maximally benefit from Bit-Column-Serial Computation (BCSeC), we design a supporting neural processing unit (NPU) architecture, named \emph{BitWave}. The challenges in the design of this NPU are as follows: C1) enabling low-overhead index parsing and bit-serial computation; C2) maintaining both wide processing array parallelism and high utilization across various workloads. In this section, we will first introduce the key modules of \emph{BitWave}, including the micro-architecture of the zero-column index parser and the compute engine with BCSeC support (Challenge 1). Then we detail the dynamic dataflow adopted in \emph{BitWave}, illustrating the efficient parallel computation with the combination of BCSeC (Challenge 2). These blocks are integrated into the overall \emph{BitWave} architecture, depicted in Figure \ref{fig:archi}.

\subsection{Zero-column Index Parser}
\label{sec:a1}
\emph{BitWave} can process compressed weights without the need for \mv{first decompressing the} data. Instead, it \mv{processes} 
bit significance information for non-zero-bit columns \mv{in the} 
Zero-column Index Parser (ZCIP). 
Figure \ref{fig:idx} illustrates the structure of the ZCIP module. Each weight index bit vector, \mv{is split into its MSB (indicating the sign bit-column of the grouped weights) and the remaining bits}. If the MSB is 1, a $Sign\_Rqst$ signal is raised, \mv{signaling the need} to read the non-zero sign bit column. Otherwise, \mv{all sign bit columns are reset to '0'.} 
The remaining index bits $Idx[6]$ to $Idx[0]$ indicate the sparsity of each data-bit column. 
\mv{This will be used to initiate} 
the \mv{appropriate} shift \mv{operations when} iterating through all the "1" bits within the zero index \mv{and serially processing the different bit positions}. 
ZCIP also derives the total number of non-zero (NZ) columns to control the number of clock cycles required to complete the current index-associated computation through $Sync. ctr$. \emph{BitWave} is capable of processing 1024 bits indexes in parallel, hence the system contains 128 parsers with all up to 8 bits wide. ZCIP is also able to be configured to dense mode. In dense mode, it generates shift control locally based on precision configuration, the overheads of indexes hence are avoided when handling deeply quantized weight.


 
 \begin{figure}[tb]
\centering
\includegraphics[width=0.78\linewidth]{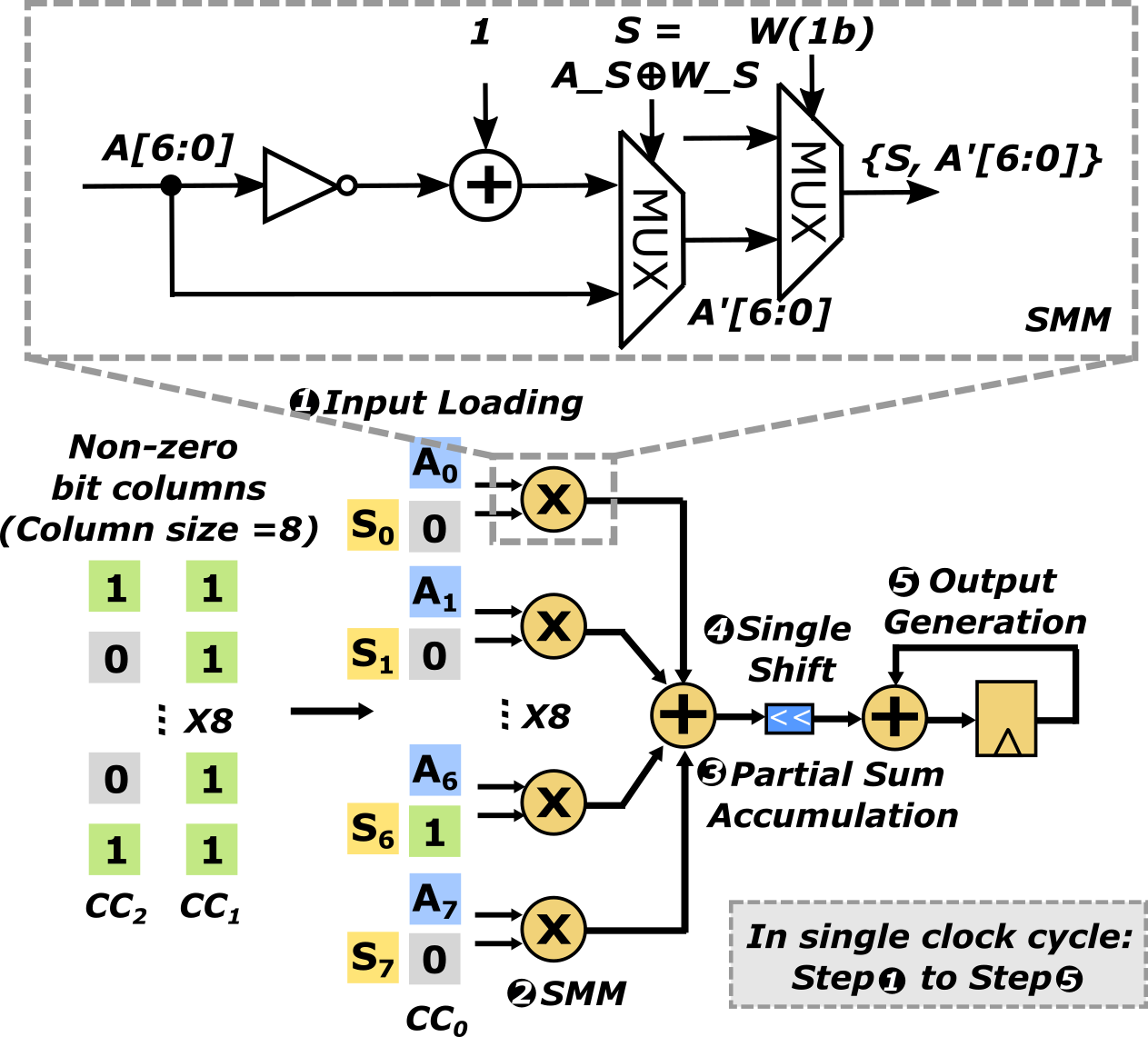}
\caption{\vs{Sign-magnitude bit-serial multiplier and BCE architecture}}
\label{fig:bce}  
\vspace{-0.4cm}
\end{figure}

\subsection{BitWave Computing Engine} 
\label{sec:a2}
The column-compressed BS compute takes place in the \emph{BitWave} Compute Engine (BCE). 
\subsubsection{Sign-magnitude multiplier}
The core of the \emph{BitWave} Compute Engine \mv{are the} sign-magnitude multipliers (SMM). In each clock cycle, BCE multiplies a single bit-column of weights with multiple full-precision two's complement represented activations (8b in the example of Figure \ref{fig:bce}). Each bit of the weight column is multiplied with a different activation using a simple AND gate.
The signs of the weights (provided by the ZCIP) and activation jointly determine the partial product's sign. If the signs are the same, the product is positive, negative otherwise. All partial products of the one-bit column are finally added together, considering their signs.

\subsubsection{BCE module}
BCE harnesses the shared-significance property in BCSeC by performing addition operations on the partial results of each bit within a column before applying a shift operation. However, this is only allowed when spatially unrolling the input channel (C) or kernel (FX or FY) dimensions along the bit-columns. In this work, we assume unrolling across C, often a multiple of eight, enabling us to utilize the hardware resources fully. Consequently, instead of performing individual shifts for each weight bit, BCE applies a single shift operation for the entire bit column after the addition stage, reducing overhead. \vs{As demostrated in Figure~\ref{fig:bce}},  \vs{starting from Step \ding{182} Input Loading}, BCE first receives \mvn{eight} 8b \vs{activations}, an 8$\times$1b weight bit column, and the sign bits. \vs{In Step \ding{183} SMM}, BCE performs a simple 8$\times$1b sign magnitude multiplication, followed by \vs{Step \ding{184} Partial Sum Accumulation}, accumulating the partial results of all elements within the column. Then, \vs{through Step \ding{185} Single Shift}, BCE  uses the shift information from ZCIP to apply a single shift operation on the partial sum of the non-zero bit column, aligning the bits based on their significance. Finally, \vs{in Step \ding{186} Output Generation}, the shifted results are stored in a local register for further processing or output generation. The sign bits and corresponding activation are reused for multiple clock cycles until all non-zero columns belonging to the same weight are processed, while the weight bits are updated every computing cycle. 

\subsection{Flexible BitWave dataflow}\label{sec:a3}
To fulfill the high throughput demands of DNNs, we deploy 512 BCE modules in parallel, together encompassing 4096 1b$\times$8b SMMs and ensuring the same performance as 512 8b$\times$8b bit-parallel PEs. However, this large number of SMMs poses a threat towards the under-utilization of PE arrays across workloads with diverse dimensions. 
To address this challenge, we equip the PE array with dataflow reconfiguration capabilities, to dynamically adapt its spatial unrolling dimensions to the layer characteristics. \cite{du202328nm} \vs{has \mvn{shown that}} different dataflow mappings shine across different layer shapes: an $OX_u, OY_u$ (XY) parallel SU is beneficial for wide layers, $C_u, K_u$ (CK) parallelism is good for deep layers, while $OX_u, FX_u$ (XFx) benefits layers with large kernel sizes. We analyze the PE utilization for MobileNetV2 and ResNet18 under each type of SU mapping on a 4096 1b$\times$8b PE array and a 512 8b$\times$8b PE array for the following four workload cases: early layer (wide but shallow shape, e.g., ResNet18 conv1 layer), late layer (narrow but deep shape, e.g., ResNet18 last conv layer), depthwise convolution (Dwcv, with one input channel, e.g., Dwcv1 of  MobileNetV2), and pointwise convolution (Pwcv with 1×1 size kernel, e.g., Pwcv1 of MobileNetV2).

\begin{figure}[tb]
\centering
\includegraphics[width=0.9\linewidth]{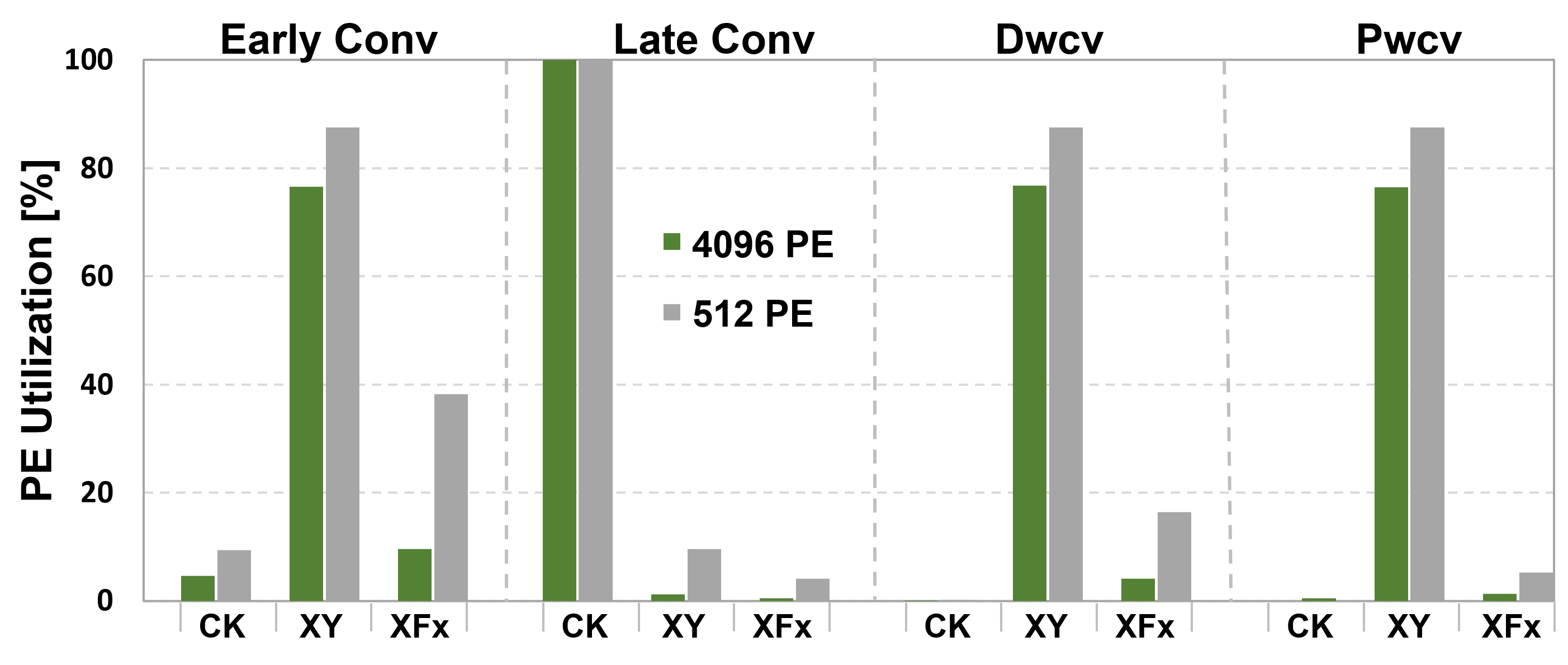}
\caption{PE utilization evaluations for a fixed SU across layer shapes} 
\label{fig:su_fx}   
\end{figure}

\begin{table}[!t]
    \caption{BitWave Spatial Unrolling and corresponding bandwidth}
    \centering
    \begin{threeparttable}
	\begin{tabular}{|c|c|c|c|c|}
		\hline
        ~ & \multirow{ 2}{*}{SUs} & W BW & Act BW   \\
		~ & ~ &  (bit/cycle)  & (bit/cycle)  \\
		\cline{1-4}
  
		$SU_1$ &$[C_u=8, OX_u=16, K_u=32] $& 256 & 1024 \\
		\cline{1-4}

		$SU_2$ &$[C_u=16, OX_u=8, K_u=32] $& 512 & 1024 \\
		\cline{1-4}

           $SU_3$ & $[C_u=32, OX_u=4, K_u=32]$  & 1024 & 1024 \\
		\cline{1-4}
           $SU_4$ & [C\_u=8, OX\_u=1, K\_u=128]  & 1024 & 64 \\
		\cline{1-4}
         $SU_5$ & $[C_u=16, OX_u=1, K_u=64]$& 1024 & 128\\
		\cline{1-4}
 $SU_6$ & $[C_u=32, OX_u=1, K_u=32]$ &  1024 & 256\\
		\cline{1-4}
		\hline
  $SU_7$ & $[G_u=64, OX_u=2, K_u=1]$\tnote{1} & 64 & 1024 \\
		\cline{1-4}
		\hline
	\end{tabular}
    \label{table:su}
    \begin{tablenotes}
        \footnotesize
        \item[1] SU specialized for the Depthwise Convolution.
      \end{tablenotes}
    \end{threeparttable}
\end{table}

It is clear from Figure~\ref{fig:su_fx} that not a single fixed SU mapping can achieve over $80\%$ utilization across all workload cases. 
It is also noteworthy that the larger-sized PE array suffers more severe under-utilization than the small-sized array, which is a disadvantage of bit-serial computation, as we discussed before.
To address these challenges, this work leverages a dynamic dataflow.
The proposed architecture can adapt the spatial parallelism along the K, C, and OX dimensions at runtime to improve utilization, while remaining compatible with the introduced BCSeC, which requires C sufficient parallelism. Specifically, the \emph{BitWave} architecture supports 7 different dataflows, summarized in Table~\ref{table:su}, which adjusts $C_u$ to compensate for a desired reduction in $OX_u/OY_u$ or $Ku$ for specific workload shapes. 
The selection of SUs as such incorporates dataflows optimized for channel-heavy convolutional layers and layers with limited input channels, depthwise layers, and fully connected layers. 
\begin{figure}[tb]
\centering
\includegraphics[width=0.98\linewidth]{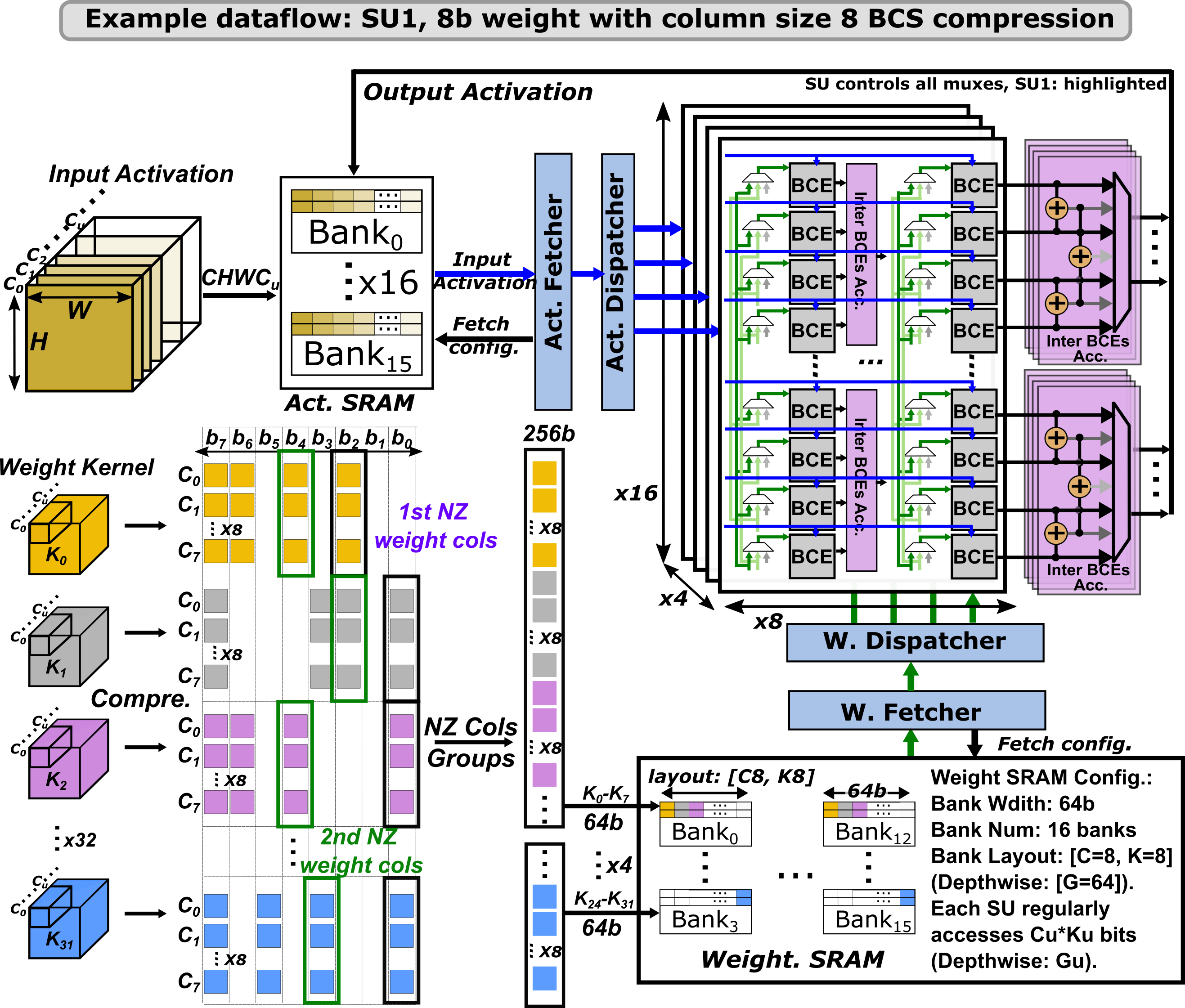}
\caption{\vs{BitWave dataflow \mvn{and memory data layout}}} 
\label{fig:df}  
\vspace{-0.3cm}
\end{figure}
 
The computational dataflow for $SU_1$ is depicted in Figure~\ref{fig:df} \vs{as an example}. Input feature maps are organized as $CHWC_u$ sequence and stored in a 16-bank activation SRAM. Weight kernels undergo offline pre-processing, preserving only the Non-Zero (NZ) bit-columns for storage. In $SU_1$, \mvn{\emph{BitWave}'s 512 BCEs are organized in a 16x32 manner} [Cu=8, OXu=16, Ku=32], \mvn{(visualized in Figure 10 in a 16x8x4 grid)}. \vs{\mvn{Under this dataflow,} \emph{BitWave} requires to access 256 bits of weights ($C_u*K_u=8*32$ bits) in each cycle, which translates to 4 banks in parallel (4 segments of 64 bits). Each 64-bit segment contains 64 same significance weight bits from 8 different input channels (the C dimension) across 8 different kernels (the K dimension). The resulting weight data layout within one bank is shown in Figure~\ref{fig:df}\mvn{(bottom left)}. \mvn{Each of the} 4 segments of 64 \mvn{weight} bits is sent to 16x8 BCEs. Each \mvn{plane of} 8x16 BCE's receives the same 1024-bit inputs, uni-casting a 64bit \mvn{input segment} to each BCE row.} 

To enable the adaptable dataflow \mvn{supporting the different SU's}, the parallelism of K, C, and OX \mvn{across the 512 BCE's can be reconfigured}, requiring parallel access to different-sized data chunks per clock cycle, as outlined in Table~\ref{table:su}. This mechanism is facilitated by the Act./W. fetcher, using the layer SU configuration. Subsequently, these data chunks are directed to the Data Dispatcher, which can be programmed to employ diverse data casting strategies for various SUs. 
\vs{It is\mvn{, however, important} to note that \emph{BitWave} ensures regular memory access \mvn{consisting of packed 64-bit segments across all possible supported SU's, and for any BCS sparisity}. The compressed weights are always directly streamed into PE array for computation \mvn{without a need for further online preprocessing. This} is both hardware-friendly and efficient, \mvn{in contrast to} methods such as EBPC \cite{cavigelli2019ebpc}, which prioritize bit-level sparsity, but compromise on memory access \mvn{regularity. For example, EBPC requires} a complex decompressor to unpack compressed data before computation, resulting in up to an additional $10.5\%$ area overhead.}

\begin{figure}[tb]
\centering
\includegraphics[width=0.75\linewidth]{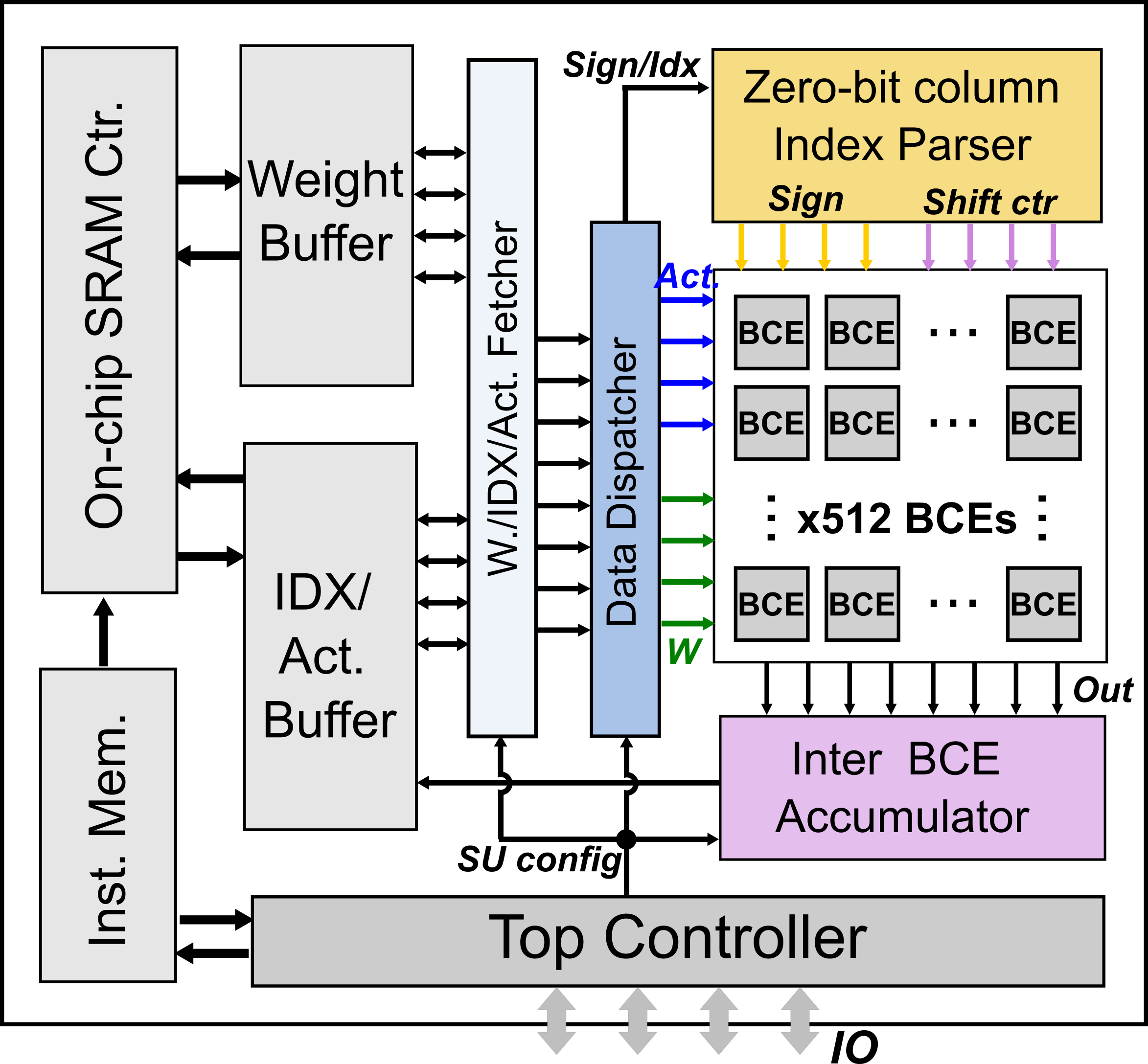}
\caption{BitWave overall architecture} 
\label{fig:archi} 
\vspace{-0.5cm}
\end{figure}

Figure \ref{fig:archi} presents the overall \emph{BitWave} architecture, featuring 512 BCEs, alongside the Zero-Column Index Parser (ZCIP), Data Fetcher, and Data Dispatcher modules. \emph{BitWave} adopts an output stationary approach and employs customized activation and weight unicasting/broadcasting methods for different SU configurations. We incorporate the design space exploration tool ZigZag \cite{mei2021zigzag} to analyze the optimal SU within support sets for each layer offline. The SU information will be stored in the instruction memory that contains all layer-specific configurations. The top controller hence plays a crucial role in dynamically adjusting spatial unrolling of the PE array, by programming the Data Fetcher and Data Dispatcher on a per-layer basis.

\section{Evaluation }\label{sec:Evaluation} 
In this section, we first explain \emph{BitWave}'s implementation details and experimental setup (Section \ref{sec:e1}). Next, we discuss the \mvn{sparsity-aware modeling of SotA} benchmarks \mvn{used for comparison} (Section \ref{sec:e2}). 
\mvn{Then, we use this implementation and models to assess the proposed optimization techniques of \emph{BitWave} in terms of energy and speedup, compared to the SotA different benchmarks (Section \ref{sec:e3}).}
Finally, Section V-D presents \mvn{\emph{BitWave}} area and power breakdown both for a single BCE and at the system level. 

\subsection{Implementation and experimental setup}\label{sec:e1}
\subsubsection{Implementation}\label{sec:e1.2}
We implement \emph{BitWave} at the RTL-level using SystemVerilog and synthesize the design in 16nm FinFET technology with Synopsys Design Compiler (DC) 
Our implementation of \emph{BitWave} comprises 512 BCEs, 256KB weight SRAM, and 256KB activation SRAM, running at a clock frequency of 250~MHz with 0.8V nominal voltage. 
The resulting netlist from the synthesis is simulated to generate a VCD (Value Changed Dump) file, which is then analyzed using DC for power assessment. We further deconstruct the synthesis reports of \emph{BitWave}'s BCE to provide detailed module-level power and area information. For DRAM energy, the DDR3 model from DRAMpower \cite{DRAMPower} is utilized.
\begin{figure*}[tb]
\centering
\includegraphics[width=1\linewidth]{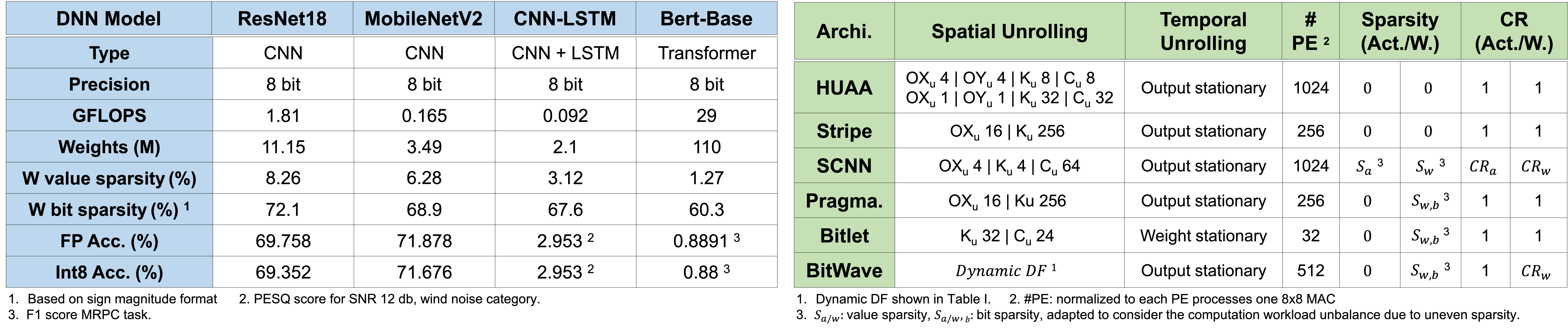}
\caption{Workload and accelerator benchmarks} 
\label{fig:accele} 
\vspace{-0.3cm}
\end{figure*}
\subsubsection{\mvn{Experimental setup}}
For the evaluation of \emph{BitWave} \mvn{and its SotA competitors}, 
we selected ResNet18 \cite{he2015deepresiduallearningimage}, MobileNetV2 \cite{sandler2018mobilenetv2}, CNN-LSTM (\cite{cnnlstm} in-house audio denoising model), and Bert-Base \cite{geetha2021improving}, which with diverse model types, GFLOPs, and parameter sizes, listed in \vs{Figure~\ref{fig:accele}(left)}.
We leverage open-source Int8 quantized networks, or - where not available - quantize the fp32 weights into Int8 precision using PyTorch's common 
post-training quantization framework. 
We provide the performance improvement breakdown by comparing the proposed optimization techniques incrementally. Starting with the introduction of the Dynamic DataFlow (DF), followed by the BCSeC exploiting bit column sparsity on sign magnitude formatted weights (DF + SM), and finally, adding the sparsity enhancement using Bit-Flip (DF + SM + BF), demonstrating the strengths of each proposed technique. 

\begin{table}[!t]
\caption{ZigZag Outputs Abbreviation definition}
\vspace*{-0.3cm}
\label{table:abbr}
\centering
\begin{tabular}{|l|l|}
\hline
 \bf{Symbol} & \bf{Meaning} \\ \hline \hline
$N_{DRAM\_read, a}$ & Number of off-chip activation DRAM read \\ \hline
$N_{DRAM\_read, w}$ & Number of off-chip weight DRAM read \\ \hline
$N_{DRAM\_writ, a}$ & Number of off-chip activation DRAM write \\ \hline
$N_{DRAM\_write, w}$ & Number of off-chip weight DRAM write \\ \hline
$N_{SRAM\_read/write}$ & Number of on-chip SRAM read/write \\
$_{-input/output/weight}$ & for input, output and weight, respectively  \\ \hline    
$N_{reg\_read/writ}$ & Number of on-chip Register read/write \\ \hline
$N_{mac}$ & The total number of MAC \\ \hline
$N_{mac, cycle}$ & The Effective number of MAC per cycle\\ \hline
\end{tabular}
\vspace*{-0.4cm}
\label{table:out}
\end{table}
\subsection{\mvn{SotA accelerator benchmarks}}
\label{sec:e2}
To further showcase the benefits of the proposed techniques, \emph{BitWave} will be compared to a broad set of accelerator baselines: (1)~Sparsity-unaware accelerators: 1a.) HUAA \cite{du202328nm}: The Hardware-Utilization-Aware Accelerator using bit-parallel, computing with dynamic dataflow reconfigurability, but without sparsity handling; 1b.) Stripes \cite{judd2016stripes}: bit-serial accelerator, without bit-level sparsity handling. (2)~Sparsity-aware accelerators: 2a.) SCNN \cite{parashar2017scnn}: a value-sparsity aware accelerator; 2b.) Pragmatic \cite{albericio2017bit}, Bitlet \cite{bitlet}: both bit-sparsity aware prototypes, where Bitlet takes advantage of weight bit sparsity and Pragmatic can leverage the bit sparsity on either weights or activations. In our comparison, weight bit sparsity is assumed for Pragmatic. 

\mvn{To make a fair comparison between \emph{BitWave} and these alternatives, all systems should be compared with an equivalent number of processing elements, and memory hierarchy. The resulting architecture assumptions for each benchmark are summarized in Figure \ref{fig:accele}(right). Due to the difficulty of re-implementing each technique at the RTL level under these constraints, the comparison is done by modeling each accelerator in a uniform way. The performance model, inspired by Sparseloop \cite{wu2023sparseloop}, is described in detail in this subsection, and validated against the RTL model of \emph{BitWave}}. 

The accelerator models are created by (STEP1) modeling each of the accelerators with their respective dataflows in an analytical accelerator modeling tool; (STEP2) deriving the relevant sparsity statistics for each workload under study; (STEP3) combining the information of STEP1 and STEP2 to extract the total number of effective operations and memory fetches for each accelerator under study; and (STEP4) determine the resulting energy and latency performance of running each workload on each accelerator. The detailed description of each step is provided as below.

\subsubsection{STEP1} Each accelerator is modeled in the ZigZag open source accelerator modeling tool \cite{mei2021zigzag} under the spatial and temporal unrolling parameters of Figure~\ref{fig:accele}(right) and a common SRAM-DRAM memory hierarchy. For each accelerator-workload combination, detailed internal operational activity counts are extracted, summarized in Table \ref{table:out}. 

\subsubsection{STEP2} In parallel, the sparsity statistics of both weights and activations are extracted for each layer of each workload at the word-level as well as the bit-level. Specifically, for each Int8 workload, the value, resp. bit-level sparsity of weights ($S_w$, resp. $S_{w,b}$) and activations ($S_a$, resp. $S_{a,b}$) is derived. It is important to note, that these sparsity statistics are adjusted to accommodate for load imbalance in the runtime scheduled accelerators.  
Secondly, for each accelerator, the weight and/or activation compression technique is taken into account, to derive the compression ratio of each tensor in the workload transferred between DRAM and/or SRAM. The compression ratio takes into account both the data reduction due to the sparse encoding, as well as the overhead due to additional required index information  
(e.g. SCNN's non-zero value coordinates, or \emph{BitWave} bit index). 
For each targeted accelerator, Figure~\ref{fig:accele}(right) summarizes the relevant sparsity and compression parameters.

\subsubsection{STEP3} 
The effective total number of MACs ($N_{mac,e}$, including zero skipping) and resulting computational clock cycles, can now be computed as: 
\begin{equation}
N_{mac,e} = N_{mac} \times (1-S_{a})  \times (1-S_{w})  
\end{equation}
\begin{equation}
CC_{mac,e} = \frac{N_{mac,e}}{N_{mac, cycle}} 
\label{cc_dram}
\end{equation}
with $N_{mac}$ the number of MACs of the workload and $N_{mac,cycle}$ the effective number of MACs per cycle for this workload when taking the accelerator's spatial unrolling into account, both analytically extracted by ZigZag (Table \ref{table:out}).

The effective total number of memory read and memory write accesses at each memory level (DRAM - SRAM - register) is derived as follows:
\begin{equation}
N_{DRAM\_read,e} = \frac{N_{DRAM\_read,w}}{CR_{w}} + \frac{N_{DRAM\_read,a}}{CR_{a}}
\label{cc_dram}
\end{equation}
and likewise, for DRAM write accesses and SRAM read/write accesses, in which all SRAM, DRAM dense access counts are again analytically extracted by ZigZag (Table \ref{table:out}). 
\subsubsection{STEP4}
The derived activity counts, can finally be combined with technology parameters expressing the unit energy per MAC operation, DRAM access, SRAM access, and register read/write. All unit costs were derived from synthesis results 
corresponding to 16 nm technology, except for the DRAM access energy, which was sourced from the open-source tool DRAMpower.





\begin{equation}
\begin{split}
Total_{Energy} &= \sum N_{(mem\_level)\_r/w,e} \times E_{unit,(mem\_level)\_r/w} \\
&\quad + N_{mac,e} \times E_{unit,mac}
\end{split}
\label{energy}
\end{equation}

For latency, it has to be considered that memory transfers can be hidden underneath compute or vice versa. Assuming that the computation initiates once all necessary data is available in the on-chip SRAM, and a separate single port SRAM for weights and I/O activations (as present in the referenced accelerator designs):
\begin{equation}
\begin{split}
Total_{cycle} = N_{DRAM\_read/write,e} + N_{SRAM\_write-output,e} \\ + max(N_{SRAM\_read-input,e}, N_{SRAM\_read-weight,e},  \\ N_{reg\_read,e}, N_{reg\_read,e}, CC_{mac,e})
\label{latency}
\end{split}
\end{equation}

The presented model has been validated against the RTL model of \emph{BitWave}, demonstrating a deviation of less than $6\%$. 
It is important to note already that the model ignores the overhead of additional complexities stemming runtime data reshuffling needed in SCNN and Bitlet. 

\begin{figure}[tb]
\centering
\includegraphics[width=0.9\linewidth]{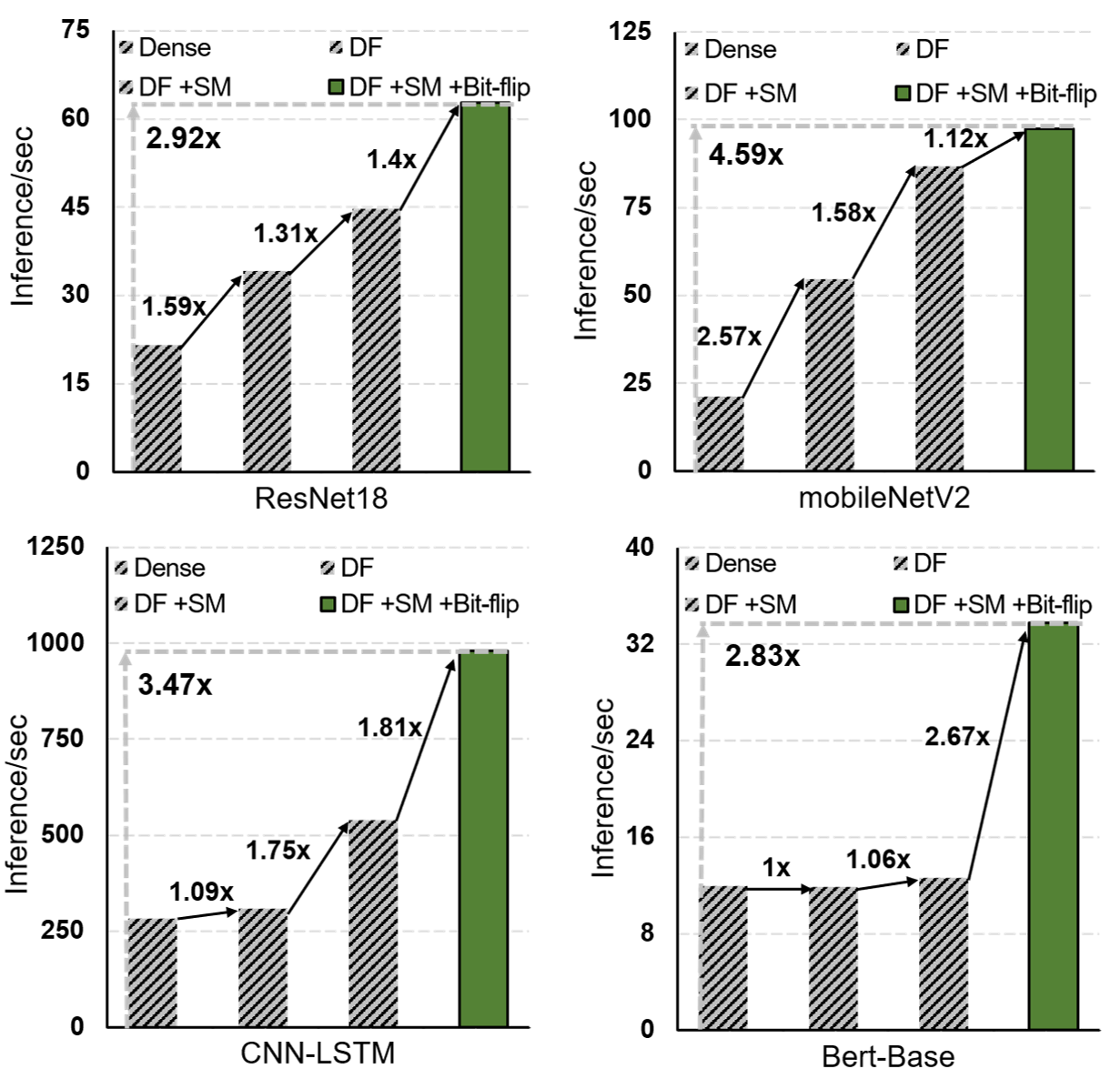}
\caption{ResNet18/MobileNet/CNN-LSTM/Bert-Base speedup breakdown. Bert-Base is running with input token size 4 (higher is better).} 
\label{fig:speed} 
\vspace{-0.5cm}
\end{figure}

\begin{figure*}[tb]
\centering
\includegraphics[width=0.8\linewidth]{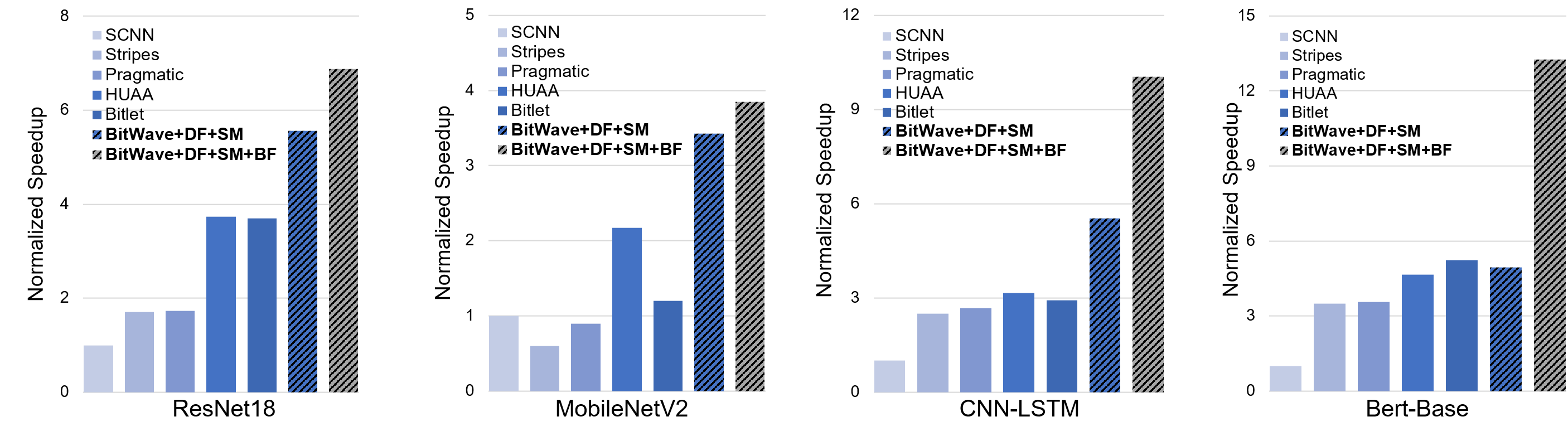}
\caption{Speedup comparison. The speedup is normalized to SCNN (higher is better).}
\label{fig:speedc}   
\vspace{-0.3cm}
\end{figure*}

\begin{figure*}[tb]
\centering
\includegraphics[width=0.8\linewidth]{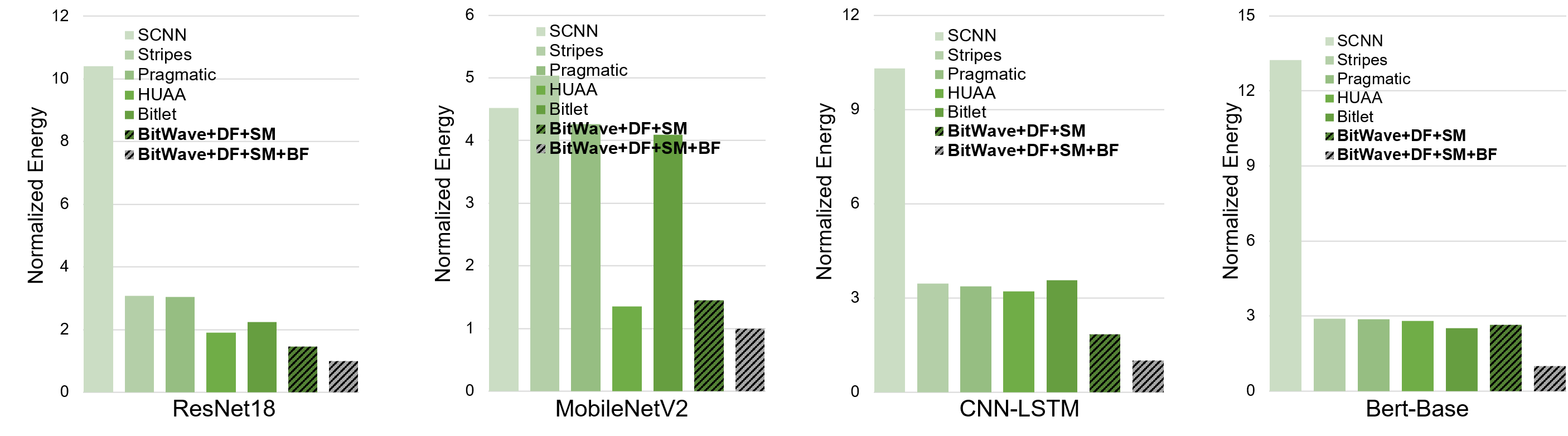}
\caption{Energy consumption comparison. The Energy consumption is normalized to \emph{BitWave}+DF+SM+BF (lower is better).} 
\label{fig:energy}   

\end{figure*}
\begin{figure}[tb]
\centering
\includegraphics[width=0.8\linewidth]{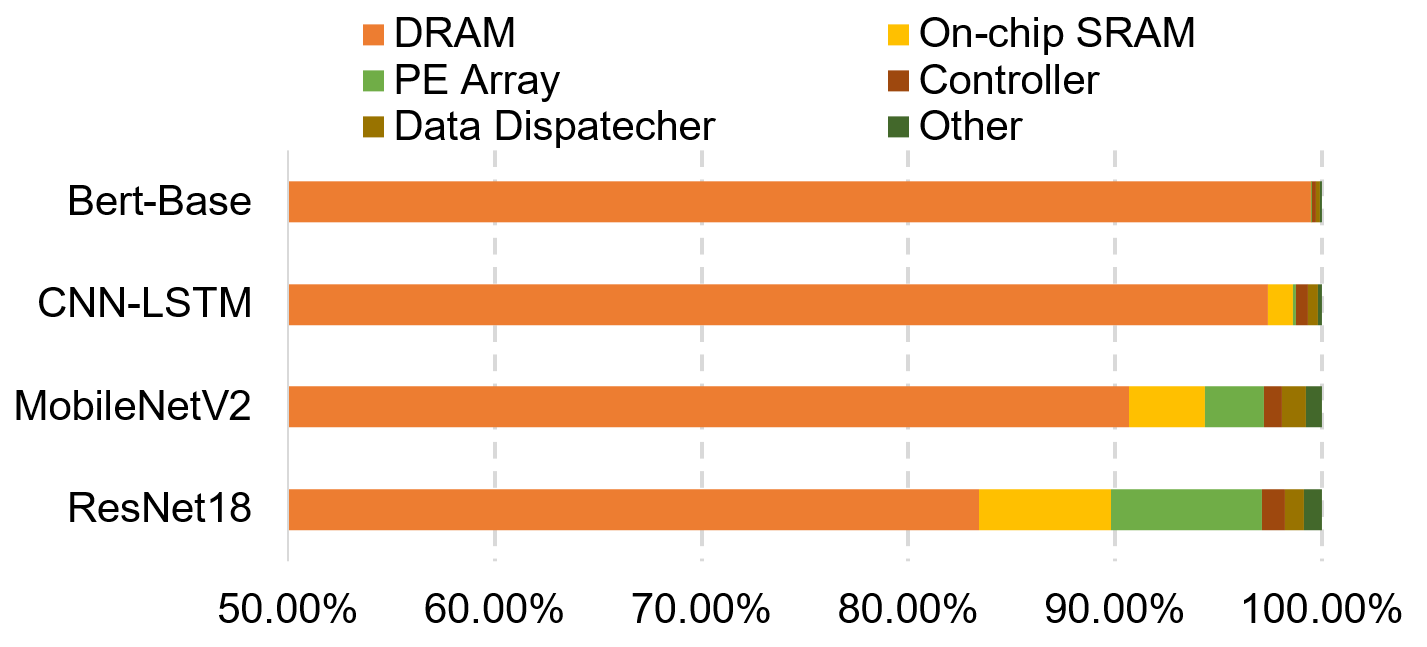}
\caption{Overall energy breakdown of BitWave and off-chip DRAM} 
\label{fig:energyb}   

\end{figure}
\begin{figure}[tb]
\centering
\includegraphics[width=1\linewidth]{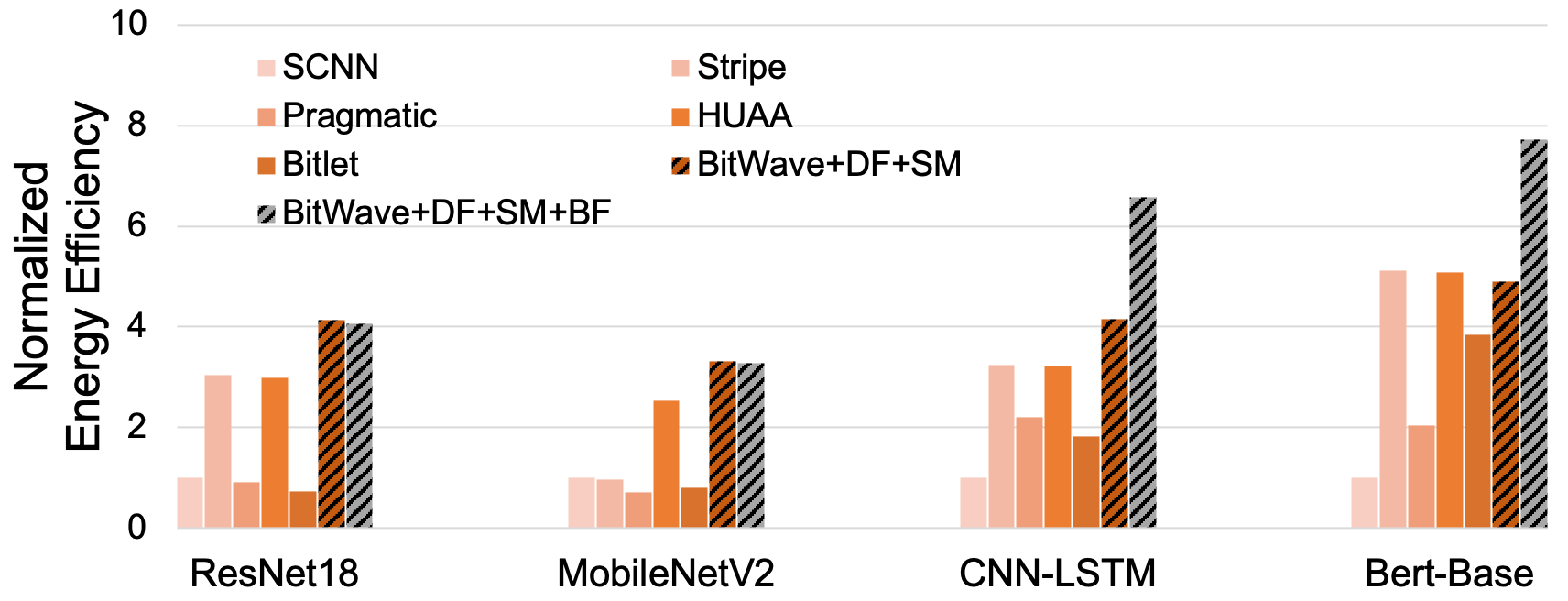}
\caption{Energy efficiency comparison. The energy efficiency is normalized to SCNN (higher is better).} 
\label{fig:ee}   
\vspace{-0.4cm}
\end{figure}
\subsection{Speedup and Energy Efficiency } \label{sec:e3}
\subsubsection{Bitwave Speedup breakdown} 
Figure~\ref{fig:speed} summarizes \emph{BitWave}'s speedup achieved by each optimization technique across the different benchmark networks. The Dense model is evaluated based on $[K_u=64, C_u=64]$, which is a common-used SU in previous works \cite{9209075, Nvidia}. \emph{BitWave}'s dynamic dataflow allows flexible spatial unrolling layer by layer, resulting in a higher PE utilization. This is clearly evident on MobileNeV2 with a $2.57\times$ performance improvement due to dataflow (DF) optimization. MobileNeV2, with its diverse layer topologies, benefits a lot from a distinct SU for each layer. Meanwhile, the CNN-LSTM and Bert-Base are less influenced by the dynamic dataflow due to their less diverse layer shapes or types. The introduction of the Sign Magnitude (SM) format-based BCSeC brings an additional speedup of $1.31\times$, $1.58\times,$ and $1.75\times$ on ResNet18, MobileNetV2, and CNN-LSTM, respectively, utilizing the abundance of zero bit columns in the original Int8 model. Although the BCSeC with only SM representation shows $1.06\times$ up on Bert-Base, adopting the Bit-Flip technique enables an additional $2.67\times$ speedup on Bert-Base with $0.5\%$ F1 score decrease but without any (re)training involvement.

\subsubsection{Speedup \mvn{vs SotA}} \label{sec:e4}
In Figure~\ref{fig:speedc}, we compare \emph{BitWave}'s performance to different SotA accelerators. All the results are normalized to "SCNN". \emph{BitWave} shows $10.1\times$ and $13.25\times$ performance gain compared to SCNN for CNN-LSTM and Bert-Base, respectively. This stems from the scarcity of value-level sparsity in the original model. SCNN's implementation of ZRE for data compression proves beneficial only under high value sparsity, demanding the involvement of specialized training techniques. When sparsity is limited, the gains from zero value skipping fall short, while the overheads of the required flexible indexing undo any performance gains, demonstrating the limitations of value sparsity. Pragmatic achieves a commendable $1.4\times$ performance boost, employing zero-bit skipping to its advantage. However, an obstacle arises in the form of workload imbalance, tempering hardware utilization. Bitlet, on the other hand, performs bit-interleaving, avoiding the complexities of bit significance alignment. But even with this feature, the computational cycle count suffers from the bit-significance teeming with non-zero bits that impede optimal hardware utilization, especially in large arrays that demand syncing across more data items. This explains why \emph{BitWave} outperforms Bitlet by over $2\times$ across all benchmarks.
HUAA boldly adopts dynamic dataflow, maintaining a high PE utilization. Yet, it fails to capitalize on the abundant bit-level sparsity, leaving potential speedup untapped. 


\subsubsection{Energy Consumption} \label{sec:e4}
Among the five evaluated architectures concerning energy consumption (Figure~\ref{fig:energy}), SCNN emerges as the least favorable option, particularly when dealing with weight-intensive networks. This is mainly attributed to the SCNN compression approach based on value-sparsity, which significantly increases both on-chip and off-chip memory accesses due to an abundance of index overheads. For example, executing Bert-Base using SCNN incurs a staggering $13.23\times$ higher energy consumption compared to \emph{BitWave}. Sparsity-aware architectures such as Stripe, Pragmatic, Bitlet, and SCNN experience $5.04\times$, $4.26\times$, $4.09\times$, and $4.51\times$ higher energy costs, respectively, especially evident in the context of MobileNetv2. This discrepancy can be attributed to their lower PE utilization under one fixed dataflow, resulting in reduced spatial data reuse and an increased need for on-chip data accesses, further exacerbating energy consumption. Although HUAA can apply dataflow flexibly, \emph{BitWave} still achieves a $2.41\times$ average energy decrease across all benchmarks with the employment of BCSeC.
\emph{BitWave}, on the other hand, can exploit both energy reduction through weight compression and the "shift then add" bit-column serial multiplier, optimizing memory footprints, increased data reuse and efficient MAC operations. The energy breakdown of \emph{BitWave}, including external DRAM memory accesses, is shown in Figure~\ref{fig:energyb}. DRAM energy is the dominant factor, especially for weight-intensive networks, where all weights must be loaded to on-chip SRAM at least once. 


\begin{table*}[!t]
    \caption{\vs{Comparison with State-of-the-Art}} 
     \begin{threeparttable}
    \centering
	\begin{tabular}{|c|c|c|c|c|c|c|c|c|}
		\hline
          ~&\multirow{ 2}{*}{\textbf{Tegra X2}}&\multirow{ 2}{*}{\textbf{GPU A100}} & \textbf{Stripe} & \textbf{Pragmatic} & \textbf{SCNN} & \textbf{Bitlet}   & \textbf{HUAA} & \textbf{BitWave} \\
           ~ & ~ &~ & \cite{judd2016stripes}& \cite{albericio2017bit} &\cite{parashar2017scnn}&  \cite{bitlet} &  \cite{du202328nm} &  \textbf{(ours)}   \\
		\cline{1-9}
         \hline 
         \hline 
		 \textbf{Technology}&16nm &7nm &65nm & 65nm& 16nm &28nm &28nm & \textbf{16nm} \\
		\cline{1-9}
              \hline 
		 \textbf{Frequency (MHz)}&1465 & 1410 & 980& - & 1000 &1000 &100-500 & \textbf{250} \\
		\cline{1-9}
              \hline     
              \textbf{Voltage (V)}&0.8 & 1.1 & - & - & - & - &0.66-1.3 &\textbf{0.8} \\
		\cline{1-9}
         \hline  
         \textbf{Sparsity support}& Input & A value & - & W/A bit & W/A value &W. bit &- & \textbf{W. bit} \\
	\cline{1-9}
              \hline  
            \textbf{Power}& 15W & 400W & - &51.6W &- &366mW&17mW-174mW & \textbf{17.56mW} \tnote{*} \\
		\cline{1-9}
              \hline    
             \textbf{Peak Performance}&  750.1 (fp32) &1248 (8b) &  \multirow{ 2}{*}{-} &  \multirow{ 2}{*}{-} &  \multirow{ 2}{*}{2000} & 372.35 (16b)& \multirow{ 2}{*}{-}& \multirow{ 2}{*}{\textbf{215.6 (8b) \tnote{*}}} \\
                 \textbf{(GOPS)} & 1330 (fp16) &2496 (4b) & ~ &~ & ~& 744.7 (8b) & ~ & ~ \\
              \hline 
              
             \textbf{Energy Efficiency}&  0.05 (fp32) &1.5-3.1 (8b) &  \multirow{ 2}{*}{-} &  \multirow{ 2}{*}{-} &  \multirow{ 2}{*}{-} & 0.667 (16b)& \multirow{ 2}{*}{7.5-11.2} & \multirow{ 2}{*}{\textbf{12.21 (8b) \tnote{*}}}\\
                 \textbf{(TOPS/W)} & 0.088 (fp16) &3.1-6.2 (4b) & ~ &~ & ~& 1.33 (8b) & ~ & ~  \\
              \hline

               \textbf{Area (mm2)}& - & 826 & 122.1 & 157 &7.9 &1.54 &7.81 & \textbf{1.138} \\
               \hline  
              \textbf{Normalized Energy}&  0.042 (fp32) &1.04-2.15 (8b) &  \multirow{ 2}{*}{-} &  \multirow{ 2}{*}{-} &  \multirow{ 2}{*}{-} & 0.667 (16b)& \multirow{ 2}{*}{7.5-11.2} & \multirow{ 2}{*}{\textbf{10.3 (8b) \tnote{*}}}\\
                 \textbf{Efficiency (TOPS/W) $\dagger$} & 0.061 (fp16) &2.15-4.31 (4b) & ~ &~ & ~& 1.33 (8b) & ~ & ~  \\
              \hline  
              \textbf{Normalized Area (mm2)  $\dagger$}& - & 13216 & 22.6 & 29.1 &24.2 &1.54 &7.81 & \textbf{3.49} \\
               \hline
              \textbf{Normalized Area}&  \multirow{ 2}{*}{-} &0.079-0.163 (8b) &  \multirow{ 2}{*}{-} &  \multirow{ 2}{*}{-} &  \multirow{ 2}{*}{-} & 433.1 (16b)& \multirow{ 2}{*}{960.3-1434.1} & \multirow{ 2}{*}{\textbf{2951.3 (8b) \tnote{*}}}\\
                 \textbf{Effici. (GOPS/W/mm2) $\dagger$} & ~ &0.163-0.326 (4b) & ~ &~ & ~& 863.6 (8b) & ~ & ~  \\
              \hline 
                
	\end{tabular}
      \label{table:sota}
    
    \end{threeparttable}
      \vspace{-0.3cm}
\end{table*}

\subsubsection{Energy Efficiency Comparison} \label{sec:e5}
Figure~\ref{fig:ee} illustrates the energy efficiency comparisons among various benchmarks and architecture baselines, with normalization to SCNN's energy efficiency. The energy efficiency values are computed based on the actual useful operations obtained from performance analysis and the corresponding energy consumption for each benchmark.
\emph{BitWave} outperforms all other architectures. For example, when handling Bert-Base, \emph{BitWave} (with all optimization techniques) achieves $7.71\times$ improvement in inference efficiency compared to SCNN, or $2.04\times$ higher energy efficiency compared to HUAA.


%
		

\subsection{Area and Power Breakdown}
\label{sec:e6}
\begin{figure}[tb]
\centering
\includegraphics[width=0.8\linewidth]{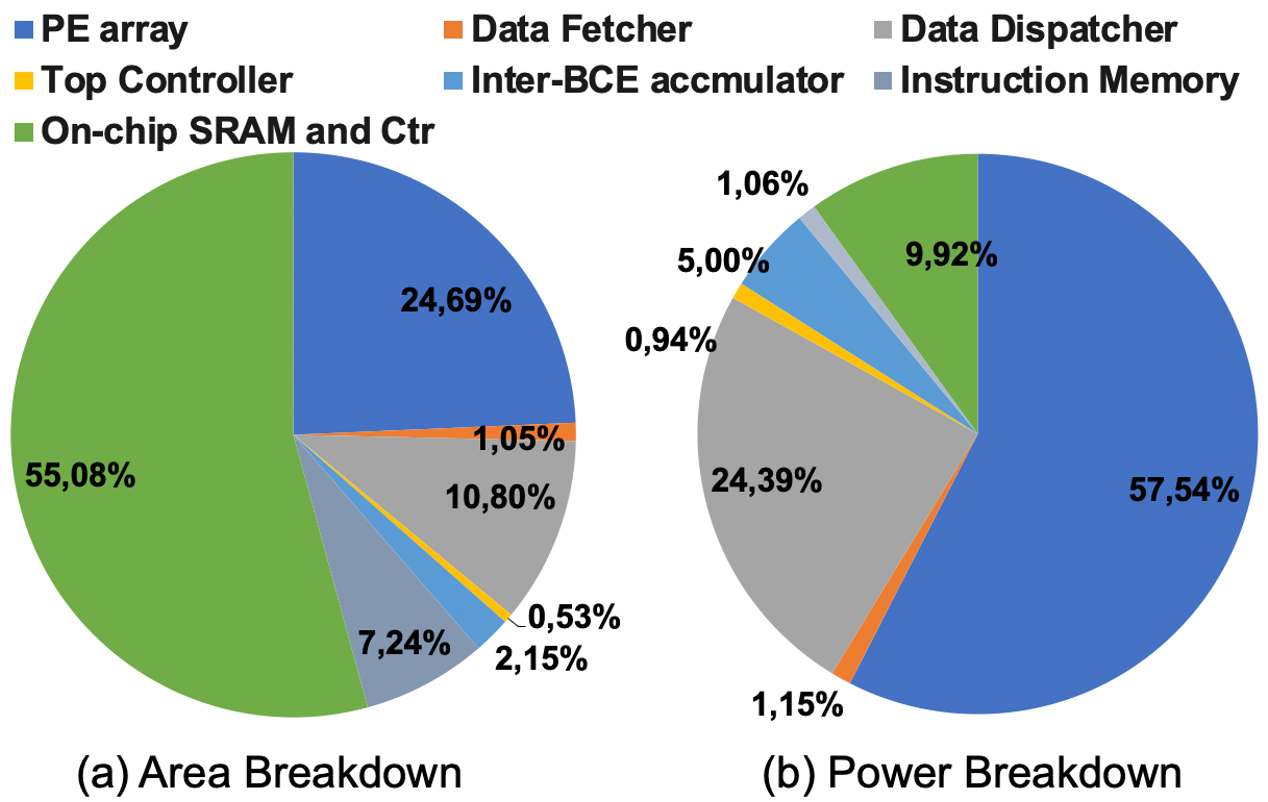}
\caption{\vs{Area and Power Breakdown of BitWave}} 
\label{fig:bk}   
\vspace{-0.4cm}
\end{figure}

The \emph{BitWave} architecture implemented in the 16nm FinFET technology node occupies an area of $1.138~mm^2$ and consumes 17.56 mW when running ResNet18 at 250MHz, 0.8V.
Figure~\ref{fig:bk} illustrates the breakdown of area and on-chip power consumption of \emph{BitWave}. The total 512KB on-chip SRAM accounts for the most significant portion of the area, occupying $55.08\%$ of the total area. On the other hand, the PE array consumes the largest portion of power, constituting \vs{$57.6\%$} of the total power, while imposing a \vs{$24.7\%$} area cost.
The Data Dispatcher, responsible for supporting the dynamic dataflow, requires high flexibility and numerous registers to serve all inputs of the PE array. This flexibility comes at the cost of $10.8\%$ and $24.4\%$ of the total area and power, respectively. Despite the area and power costs associated with the flexible data Dispatcher in \emph{BitWave}, it can demonstrate a higher energy efficiency at the system level, as discussed before. 

\begin{table}[!t]
    \caption{Comparison of Area and Power between three types PE}
    \centering
	\begin{tabular}{|c|c|c|c|}
		\hline
		\multirow{ 2}{*}{PE Type} & Power & Area  \\
		~ & (mW) & (um$^2$) \\
		\cline{1-3}
		\hline
		One 8x8 Bit-parallel PE &  2.13e-02 & 98.029   \\
		\cline{1-3}
          \hline
		Eight 1x8 Bit-serial PE &  5.71e-02 &443.284  \\
		\cline{1-3}
           \hline
            Eight 1x8 Bit-column-serial PE &  1.71e-02 & 123.431  \\
		\cline{1-3}
		\hline
	\end{tabular}
    \label{table:pe}
    \vspace{-0.4cm}
\end{table}

To comprehensively compare the processing elements (PEs) used in \emph{BitWave}, three types of PEs are evaluated: bit-parallel, bit-serial, and our bit-column-serial PE, as shown in Table~\ref{table:pe}. These PEs are designed for performing 8x8 multiplications, each handling the task differently. The bit-parallel PE exhibits a compact architecture and consumes the least area among the three types. However, it is obviously less effective in utilizing bit-level sparsity. On the other hand, the traditional bit-serial PE can leverage bit-level sparsity more efficiently, but this comes at the expense of increased area and power consumption, as each multiplier requires additional resources like shifters and registers for partial product accumulation. Our bit-column-serial PE strikes a balance between area and energy efficiency. Although it has a $1.26\times$ area overhead compared to the bit-parallel PE, it costs an average of $1.25\times$ less power through its add-then-shift structure, providing a promising solution for optimizing tasks involving bit-level sparsity. 


 
\section{Related Work \mv{and SotA comparison}} \label{sec:sota}
Table \ref{table:sota} summarizes the state-of-the-art (SOTA) specifications. \vs{To conduct a fair comparison, the normalized energy efficiency and area, as well as area efficiency to 28nm technology are provided as well.}
\emph{BitWave} is the only architecture combining extreme dataflow flexibility with advanced sparsity handling. \vs{Hence, it is capable of} reducing memory accesses and skipping redundant computations simultaneously, while maintaining high PE utilization throughout the network.
This results in \emph{BitWave} achieving the \vs{best area efficiency} across all benchmarks \vs{at the same technology node.}  Unlike previous approaches that utilized bit-level \cite{delmas2019bit, bitlet, sharify2019laconic, albericio2017bit, yang2021fusekna} or value-level sparsity \cite{li2022ristretto, 10071080, gondimalla2019sparten, parashar2017scnn} to remove redundant computations or employ costly methods to compress irregular zero-values or zero-bits in weights or activations, \emph{BitWave} focuses on the inherent patterned bit-column sparsity with the assistance of sign-magnitude formatted data \vs{in a hardware-friendly way}. Moreover, while previous works \cite{zhao2020bitpruner, sun2023bit, li2022bitcluster} required (re)training processes to balance the workload for higher hardware utilization, \emph{BitWave} leverages the "one-shot" (re)training-free Bit-Flip to achieve further performance boost. 



\section{Conclusion}
This study introduced an innovative approach called bit-column-serial computation (BCSeC), which was incorporated into a compatible hardware design named \emph{BitWave}. \emph{BitWave} efficiently reduces redundant computation and memory access by leveraging structured bit-level sparsity in sign-magnitude represented weights, including a "shift then add" computation unit, further enhancing energy efficiency. Combining dynamic dataflow and the Bit-Flip method with BCSeC, \emph{BitWave} achieved an impressive performance increase. Specifically, it outperforms Stripe, Pragmatic, SCNN, Bitlet, and HUAA with $4.7\times$, $4.5\times$, $13.25\times$, $4.1\times$, and $3.43\times$ higher performance, respectively. Moreover, \emph{BitWave} demonstrates remarkable gains in energy efficiency, with maximally $3.36\times$, $4.63\times$, $7.71\times$, $5.53\times$, and $2.04\times$ boost when running the same benchmarks.





\section*{Acknowledgements}
This project has been partly funded by the European Research Council (ERC) under grant agreement No.101088865, the Flanders AI Research Program and KU Leuven, long-term structural Methusalem funding by the Flemish Government.


\bibliographystyle{IEEEtranS}
\bibliography{refs}

\begin{thebibliography}{10}
\providecommand{\url}[1]{#1}
\csname url@samestyle\endcsname
\providecommand{\newblock}{\relax}
\providecommand{\bibinfo}[2]{#2}
\providecommand{\BIBentrySTDinterwordspacing}{\spaceskip=0pt\relax}
\providecommand{\BIBentryALTinterwordstretchfactor}{4}
\providecommand{\BIBentryALTinterwordspacing}{\spaceskip=\fontdimen2\font plus
\BIBentryALTinterwordstretchfactor\fontdimen3\font minus \fontdimen4\font\relax}
\providecommand{\BIBforeignlanguage}[2]{{%
\expandafter\ifx\csname l@#1\endcsname\relax
\typeout{** WARNING: IEEEtranS.bst: No hyphenation pattern has been}%
\typeout{** loaded for the language `#1'. Using the pattern for}%
\typeout{** the default language instead.}%
\else
\language=\csname l@#1\endcsname
\fi
#2}}
\providecommand{\BIBdecl}{\relax}
\BIBdecl

\bibitem{Nvidia}
``Nvidia, “nvidia deep learning accelerator,” 2018. [online], available: http://nvdla.org/primer.html.''

\bibitem{albericio2017bit}
J.~Albericio, A.~Delm{\'a}s, P.~Judd, S.~Sharify, G.~O'Leary, R.~Genov, and A.~Moshovos, ``Bit-pragmatic deep neural network computing,'' in \emph{Proceedings of the 50th Annual IEEE/ACM International Symposium on Microarchitecture}, 2017, pp. 382--394.

\bibitem{an202329}
H.~An, Y.~Chen, Z.~Fan, Q.~Zhang, P.~Abillama, H.-S. Kim, D.~Blaauw, and D.~Sylvester, ``29.3 an 8.09 tops/w neural engine leveraging bit-sparsified sign-magnitude multiplications and dual adder trees,'' in \emph{2023 IEEE International Solid-State Circuits Conference (ISSCC)}.\hskip 1em plus 0.5em minus 0.4em\relax IEEE, 2023, pp. 422--424.

\bibitem{cavigelli2019ebpc}
L.~Cavigelli, G.~Rutishauser, and L.~Benini, ``Ebpc: Extended bit-plane compression for deep neural network inference and training accelerators,'' \emph{IEEE Journal on Emerging and Selected Topics in Circuits and Systems}, vol.~9, no.~4, pp. 723--734, 2019.

\bibitem{choi2016towards}
Y.~Choi, M.~El-Khamy, and J.~Lee, ``Towards the limit of network quantization,'' \emph{arXiv preprint arXiv:1612.01543}, 2016.

\bibitem{cnnlstm}
B.~DeFraene, ``Private communication for inference scripts and computing pesq/stoi scores,'' \emph{NXP Semiconductors, Belgium}.

\bibitem{delmas2019bit}
A.~Delmas~Lascorz, P.~Judd, D.~M. Stuart, Z.~Poulos, M.~Mahmoud, S.~Sharify, M.~Nikolic, K.~Siu, and A.~Moshovos, ``Bit-tactical: A software/hardware approach to exploiting value and bit sparsity in neural networks,'' in \emph{Proceedings of the Twenty-Fourth International Conference on Architectural Support for Programming Languages and Operating Systems}, 2019, pp. 749--763.

\bibitem{devlin2019bertpretrainingdeepbidirectional}
\BIBentryALTinterwordspacing
J.~Devlin, M.-W. Chang, K.~Lee, and K.~Toutanova, ``Bert: Pre-training of deep bidirectional transformers for language understanding,'' 2019. [Online]. Available: \url{https://arxiv.org/abs/1810.04805}
\BIBentrySTDinterwordspacing

\bibitem{du202328nm}
C.-Y. Du, C.-F. Tsai, W.-C. Chen, L.-Y. Lin, N.-S. Chang, C.-P. Lin, C.-S. Chen, and C.-H. Yang, ``A 28nm 11.2 tops/w hardware-utilization-aware neural-network accelerator with dynamic dataflow,'' in \emph{2023 IEEE International Solid-State Circuits Conference (ISSCC)}.\hskip 1em plus 0.5em minus 0.4em\relax IEEE, 2023, pp. 1--3.

\bibitem{geetha2021improving}
M.~Geetha and D.~K. Renuka, ``Improving the performance of aspect based sentiment analysis using fine-tuned bert base uncased model,'' \emph{International Journal of Intelligent Networks}, vol.~2, pp. 64--69, 2021.

\bibitem{gholami2022survey}
A.~Gholami, S.~Kim, Z.~Dong, Z.~Yao, M.~W. Mahoney, and K.~Keutzer, ``A survey of quantization methods for efficient neural network inference,'' in \emph{Low-Power Computer Vision}.\hskip 1em plus 0.5em minus 0.4em\relax Chapman and Hall/CRC, 2022, pp. 291--326.

\bibitem{gondimalla2019sparten}
A.~Gondimalla, N.~Chesnut, M.~Thottethodi, and T.~Vijaykumar, ``Sparten: A sparse tensor accelerator for convolutional neural networks,'' in \emph{Proceedings of the 52nd Annual IEEE/ACM International Symposium on Microarchitecture}, 2019, pp. 151--165.

\bibitem{han2015deep}
S.~Han, H.~Mao, and W.~J. Dally, ``Deep compression: Compressing deep neural networks with pruning, trained quantization and huffman coding,'' \emph{arXiv preprint arXiv:1510.00149}, 2015.

\bibitem{he2015deepresiduallearningimage}
\BIBentryALTinterwordspacing
K.~He, X.~Zhang, S.~Ren, and J.~Sun, ``Deep residual learning for image recognition,'' 2015. [Online]. Available: \url{https://arxiv.org/abs/1512.03385}
\BIBentrySTDinterwordspacing

\bibitem{judd2016stripes}
P.~Judd, J.~Albericio, T.~Hetherington, T.~M. Aamodt, and A.~Moshovos, ``Stripes: Bit-serial deep neural network computing,'' in \emph{2016 49th Annual IEEE/ACM International Symposium on Microarchitecture (MICRO)}.\hskip 1em plus 0.5em minus 0.4em\relax IEEE, 2016, pp. 1--12.

\bibitem{DRAMPower}
Y.~L. Karthik~Chandrasekar, Christian~Weis, ``Open-source dram power and energy estimation tool,'' in \emph{URL: http://www.drampower.info}.

\bibitem{9076333}
H.~Kwon, P.~Chatarasi, V.~Sarkar, T.~Krishna, M.~Pellauer, and A.~Parashar, ``Maestro: A data-centric approach to understand reuse, performance, and hardware cost of dnn mappings,'' \emph{IEEE Micro}, vol.~40, no.~3, pp. 20--29, 2020.

\bibitem{9407116}
H.~Kwon, L.~Lai, M.~Pellauer, T.~Krishna, Y.-H. Chen, and V.~Chandra, ``Heterogeneous dataflow accelerators for multi-dnn workloads,'' in \emph{2021 IEEE International Symposium on High-Performance Computer Architecture (HPCA)}, 2021, pp. 71--83.

\bibitem{li2022bitcluster}
A.~Li, H.~Mo, W.~Zhu, Q.~Li, S.~Yin, S.~Wei, and L.~Liu, ``Bitcluster: Fine-grained weight quantization for load-balanced bit-serial neural network accelerators,'' \emph{IEEE Transactions on Computer-Aided Design of Integrated Circuits and Systems}, vol.~41, no.~11, pp. 4747--4757, 2022.

\bibitem{li2022ristretto}
G.~Li, W.~Xu, Z.~Song, N.~Jing, J.~Cheng, and X.~Liang, ``Ristretto: An atomized processing architecture for sparsity-condensed stream flow in cnn,'' in \emph{2022 55th IEEE/ACM International Symposium on Microarchitecture (MICRO)}.\hskip 1em plus 0.5em minus 0.4em\relax IEEE, 2022, pp. 1434--1450.

\bibitem{liang2021pruning}
T.~Liang, J.~Glossner, L.~Wang, S.~Shi, and X.~Zhang, ``Pruning and quantization for deep neural network acceleration: A survey,'' \emph{Neurocomputing}, vol. 461, pp. 370--403, 2021.

\bibitem{liu2022s2ta}
Z.-G. Liu, P.~N. Whatmough, Y.~Zhu, and M.~Mattina, ``S2ta: Exploiting structured sparsity for energy-efficient mobile cnn acceleration,'' in \emph{2022 IEEE International Symposium on High-Performance Computer Architecture (HPCA)}.\hskip 1em plus 0.5em minus 0.4em\relax IEEE, 2022, pp. 573--586.

\bibitem{bitlet}
H.~Lu, L.~Chang, C.~Li, Z.~Zhu, S.~Lu, Y.~Liu, and M.~Zhang, ``Distilling bit-level sparsity parallelism for general purpose deep learning acceleration,'' in \emph{MICRO-54: 54th Annual IEEE/ACM International Symposium on Microarchitecture}, 2021, pp. 963--976.

\bibitem{mei2021zigzag}
L.~Mei, P.~Houshmand, V.~Jain, S.~Giraldo, and M.~Verhelst, ``Zigzag: Enlarging joint architecture-mapping design space exploration for dnn accelerators,'' \emph{IEEE Transactions on Computers}, vol.~70, no.~8, pp. 1160--1174, 2021.

\bibitem{parashar2017scnn}
A.~Parashar, M.~Rhu, A.~Mukkara, A.~Puglielli, R.~Venkatesan, B.~Khailany, J.~Emer, S.~W. Keckler, and W.~J. Dally, ``Scnn: An accelerator for compressed-sparse convolutional neural networks,'' \emph{ACM SIGARCH computer architecture news}, vol.~45, no.~2, pp. 27--40, 2017.

\bibitem{rix2001perceptual}
A.~W. Rix, J.~G. Beerends, M.~P. Hollier, and A.~P. Hekstra, ``Perceptual evaluation of speech quality (pesq)-a new method for speech quality assessment of telephone networks and codecs,'' in \emph{2001 IEEE international conference on acoustics, speech, and signal processing. Proceedings (Cat. No. 01CH37221)}, vol.~2.\hskip 1em plus 0.5em minus 0.4em\relax IEEE, 2001, pp. 749--752.

\bibitem{sandler2018mobilenetv2}
M.~Sandler, A.~Howard, M.~Zhu, A.~Zhmoginov, and L.-C. Chen, ``Mobilenetv2: Inverted residuals and linear bottlenecks,'' in \emph{Proceedings of the IEEE conference on computer vision and pattern recognition}, 2018, pp. 4510--4520.

\bibitem{sharify2019laconic}
S.~Sharify, A.~D. Lascorz, M.~Mahmoud, M.~Nikolic, K.~Siu, D.~M. Stuart, Z.~Poulos, and A.~Moshovos, ``Laconic deep learning inference acceleration,'' in \emph{Proceedings of the 46th International Symposium on Computer Architecture}, 2019, pp. 304--317.

\bibitem{10129330}
M.~Shi, S.~Colleman, C.~VanDeMieroop, A.~Joseph, M.~Meijer, W.~Dehaene, and M.~Verhelst, ``Cmds: Cross-layer dataflow optimization for dnn accelerators exploiting multi-bank memories,'' in \emph{2023 24th International Symposium on Quality Electronic Design (ISQED)}, 2023, pp. 1--8.

\bibitem{sun2023bit}
W.~Sun, Z.~Zou, D.~Liu, W.~Sun, S.~Chen, and Y.~Kang, ``Bit-balance: Model-hardware co-design for accelerating nns by exploiting bit-level sparsity,'' \emph{arXiv preprint arXiv:2302.00201}, 2023.

\bibitem{tang2022mixed}
C.~Tang, K.~Ouyang, Z.~Wang, Y.~Zhu, W.~Ji, Y.~Wang, and W.~Zhu, ``Mixed-precision neural network quantization via learned layer-wise importance,'' in \emph{European Conference on Computer Vision}.\hskip 1em plus 0.5em minus 0.4em\relax Springer, 2022, pp. 259--275.

\bibitem{9209075}
F.~Tu, W.~Wu, Y.~Wang, H.~Chen, F.~Xiong, M.~Shi, N.~Li, J.~Deng, T.~Chen, L.~Liu, S.~Wei, Y.~Xie, and S.~Yin, ``Evolver: A deep learning processor with on-device quantization–voltage–frequency tuning,'' \emph{IEEE Journal of Solid-State Circuits}, vol.~56, no.~2, pp. 658--673, 2021.

\bibitem{9950755}
M.~Verhelst, M.~Shi, and L.~Mei, ``Ml processors are going multi-core: A performance dream or a scheduling nightmare?'' \emph{IEEE Solid-State Circuits Magazine}, vol.~14, no.~4, pp. 18--27, 2022.

\bibitem{wei2023qdrop}
X.~Wei, R.~Gong, Y.~Li, X.~Liu, and F.~Yu, ``Qdrop: Randomly dropping quantization for extremely low-bit post-training quantization,'' 2023.

\bibitem{wen2016learning}
W.~Wen, C.~Wu, Y.~Wang, Y.~Chen, and H.~Li, ``Learning structured sparsity in deep neural networks,'' \emph{Advances in neural information processing systems}, vol.~29, 2016.

\bibitem{wu2023sparseloop}
Y.~N. Wu, P.-A. Tsai, A.~Parashar, V.~Sze, and J.~S. Emer, ``Sparseloop: An analytical approach to sparse tensor accelerator modeling,'' 2023.

\bibitem{Xia_2023_CVPR}
B.~Xia, J.~He, Y.~Zhang, Y.~Wang, Y.~Tian, W.~Yang, and L.~Van~Gool, ``Structured sparsity learning for efficient video super-resolution,'' in \emph{Proceedings of the IEEE/CVF Conference on Computer Vision and Pattern Recognition (CVPR)}, June 2023, pp. 22\,638--22\,647.

\bibitem{xiao2023smoothquant}
G.~Xiao, J.~Lin, M.~Seznec, H.~Wu, J.~Demouth, and S.~Han, ``Smoothquant: Accurate and efficient post-training quantization for large language models,'' in \emph{International Conference on Machine Learning}.\hskip 1em plus 0.5em minus 0.4em\relax PMLR, 2023, pp. 38\,087--38\,099.

\bibitem{yang2021fusekna}
J.~Yang, Z.~Zhang, Z.~Liu, J.~Zhou, L.~Liu, S.~Wei, and S.~Yin, ``Fusekna: Fused kernel convolution based accelerator for deep neural networks,'' in \emph{2021 IEEE International Symposium on High-Performance Computer Architecture (HPCA)}.\hskip 1em plus 0.5em minus 0.4em\relax IEEE, 2021, pp. 894--907.

\bibitem{yang2019quantization}
J.~Yang, X.~Shen, J.~Xing, X.~Tian, H.~Li, B.~Deng, J.~Huang, and X.-s. Hua, ``Quantization networks,'' in \emph{Proceedings of the IEEE/CVF Conference on Computer Vision and Pattern Recognition}, 2019, pp. 7308--7316.

\bibitem{10071080}
Y.~Yang, J.~S. Emer, and D.~Sanchez, ``Isosceles: Accelerating sparse cnns through inter-layer pipelining,'' in \emph{2023 IEEE International Symposium on High-Performance Computer Architecture (HPCA)}, 2023, pp. 598--610.

\bibitem{yao2022zeroquant}
Z.~Yao, R.~Yazdani~Aminabadi, M.~Zhang, X.~Wu, C.~Li, and Y.~He, ``Zeroquant: Efficient and affordable post-training quantization for large-scale transformers,'' \emph{Advances in Neural Information Processing Systems}, vol.~35, pp. 27\,168--27\,183, 2022.

\bibitem{zhang2016cambricon}
S.~Zhang, Z.~Du, L.~Zhang, H.~Lan, S.~Liu, L.~Li, Q.~Guo, T.~Chen, and Y.~Chen, ``Cambricon-x: An accelerator for sparse neural networks,'' in \emph{2016 49th Annual IEEE/ACM International Symposium on Microarchitecture (MICRO)}.\hskip 1em plus 0.5em minus 0.4em\relax IEEE, 2016, pp. 1--12.

\bibitem{zhao2020bitpruner}
X.~Zhao, Y.~Wang, C.~Liu, C.~Shi, K.~Tu, and L.~Zhang, ``Bitpruner: Network pruning for bit-serial accelerators,'' in \emph{2020 57th ACM/IEEE Design Automation Conference (DAC)}.\hskip 1em plus 0.5em minus 0.4em\relax IEEE, 2020, pp. 1--6.

\bibitem{zhu2017prune}
M.~Zhu and S.~Gupta, ``To prune, or not to prune: exploring the efficacy of pruning for model compression,'' \emph{arXiv preprint arXiv:1710.01878}, 2017.

\end{thebibliography}

\end{document}